\documentclass[%
 reprint,
 amsmath,amssymb,
 aps,
superscriptaddress,]{revtex4-2}

\usepackage{graphicx}
\usepackage{dcolumn}
\usepackage{bm}

\begin{document}

\preprint{APS/123-QED}

\title{High-Strength Amorphous Silicon Carbide for Nanomechanics}

\author{Minxing Xu}
\affiliation{Department of Precision and Microsystems Engineering, Delft University of Technology, Delft 2628 CD, The Netherlands}
\affiliation{Kavli Institute of Nanoscience, Department of Quantum Nanoscience, Delft University of Technology, Delft 2628 CD, The Netherlands}%
\author{Dongil Shin}
\affiliation{Department of Precision and Microsystems Engineering, Delft University of Technology, Delft 2628 CD, The Netherlands}
\affiliation{Department of Materials Science and Engineering, Delft University of Technology, Delft 2628 CD, The Netherlands}
\author{Paolo M. Sberna} 
\affiliation{Else Kooi Laboratory, Faculty of Electrical Engineering, Mathematics and Computer Science , Delft University of Technology, Delft 2628 CD, The Netherlands}
\author{Roald van der Kolk} 
\affiliation{Kavli Nanolab, Department of Quantum Nanoscience, Delft University of Technology, Delft 2628 CD, The Netherlands}
\author{Andrea Cupertino}
\affiliation{Department of Precision and Microsystems Engineering, Delft University of Technology, Delft 2628 CD, The Netherlands}
\author{Miguel A. Bessa}
\email{miguel\_bessa@brown.edu}
\affiliation{School of Engineering, Brown University, Providence, RI 02912, USA}
\author{Richard A. Norte}
\email{R.A.Norte@tudelft.nl}
\affiliation{Department of Precision and Microsystems Engineering, Delft University of Technology, Delft 2628 CD, The Netherlands}
\affiliation{Kavli Institute of Nanoscience, Department of Quantum Nanoscience, Delft University of Technology, Delft 2628 CD, The Netherlands}%

\date{\today}

\begin{abstract}
For decades, mechanical resonators with high sensitivity have been realized using thin-film materials under high tensile loads. Although there have been remarkable strides in achieving low-dissipation mechanical sensors by utilizing high tensile stress, the performance of even the best strategy is limited by the tensile fracture strength of the resonator materials. In this study, a wafer-scale amorphous thin film is uncovered, which has the highest ultimate tensile strength ever measured for a nanostructured amorphous material. This silicon carbide (SiC) material exhibits an ultimate tensile strength of over 10 GPa, reaching the regime reserved for strong crystalline materials and approaching levels experimentally shown in graphene nanoribbons. Amorphous SiC strings with high aspect ratios are fabricated, with mechanical modes exceeding quality factors $10^8$ at room temperature, the highest value achieved among SiC resonators. These performances are demonstrated faithfully after characterizing the mechanical properties of the thin film using the resonance behaviors of free-standing resonators. This robust thin-film material has significant potential for applications in nanomechanical sensors, solar cells, biological applications, space exploration and other areas requiring strength and stability in dynamic environments. The findings of this study open up new possibilities for the use of amorphous thin-film materials in high-performance applications.
\end{abstract}

\maketitle

\section{\label{sec:level1}Introduction}

Advances in nanotechnology have revolutionized various fields, with the development of tensile-loaded, thin-film mechanical devices playing a pivotal role in state-of-the-art force, acceleration, and displacement sensing \cite{Krause2012,Pratt2021,Manzaneque2022,Sanchez2012,Pate2020,halg2021membrane,reinhardt2016}. Two approaches are used to boost the sensitivity of nanomechanical resonators under tensile loads. One approach fabricates the resonators using different materials in pursuit of films with inherent high-stress and low mechanical loss tangents (i.e high mechanical quality factors). In room temperature environments, high-tensile amorphous silicon nitride (a-Si$_3$N$_4$) nanomechanical resonators have marked some of the best performing devices in ultra-sensitive mechanical detectors \cite{Verbridge2006,Norte2016,Ghadimi2018,Fedorov2020,Shin2021,Bereyhi2022}. Crystalline thin film materials (e.g. crystalline silicon (c-Si) \cite{Beccari2022}, crystalline silicon carbide (c-SiC) \cite{Romero2020,Klas2022}) and graphene are expected to have higher theoretical limits, but their projected performance relies on having perfect crystal structures with no defects (incl. edge defects). Additionally it is difficult to attain crystalline films \cite{Romero2020,Beccari2022} which can be easily deposited, have good film isotropy \cite{Manjeshwar2022} and few lattice imperfections \cite{Romero2020,Beccari2022}. 

The other approach to boost sensor performance involves innovative resonator designs that concentrate stress in key areas. These designs are constrained by the thin film materials' tensile fracture limits or ultimate tensile strength (UTS). Nanostructuring reduces the UTS due to introduced crystalline defects \cite{gu2013microstructure,zhao2022controlling}. For example, the UTS of a-Si$_3$N$_4$ thin film has been shown to be 6.8 GPa \cite{Bereyhi2019}. To date, only crystalline and 2D materials have experimentally demonstrated UTS surpassing 10 GPa after being top-down nanofabricated \cite{Kwon2015,Rasool2013,zhang2015fracture,Goldsche2018}. Among 2D crystalline materials, graphene harbors one of the highest theoretical UTS \cite{lee2008measurement,zhang2015fracture}, but practically reaching the limit is also challenging due to lattice imperfections \cite{wang2012effect}, atomically irregular edges \cite{Goldsche2018}, or sparser grain boundaries \cite{xu2018enhancing} resulting from nanostructuring processes, which lead to a reduced fracture limit when it is tensile-loaded. In this regard, amorphous thin films with high UTS offer more design freedom for free-standing nanostructures, due to their lack of both crystalline defects and sensitivity to notches \cite{qu2012metallic,jiang2021structures,greer1995metallic,telford2004case}. Apart from allowing the enhancement of the Q factor of nanomechanical resonators, higher material UTS can enable the devices to perform better in diverse and harsh vibrational environments.

Amorphous SiC (a-SiC) thin film is gaining traction due to its remarkable mechanical strength and versatile properties \cite{Wijesundara2011, Gerhardt2011, Kimoto2014}. It holds unique advantages over its crystalline counterparts, such as lower deposition temperature and adaptability to various substrates \cite{Morana2012,Iliescu2012}, enabling deposition on large wafer scales. This material stands out in applications requiring protective coatings and in the development of MEMS sensors and integrated photonics, due to its resilience to mechanical wear \cite{Blum1999} and chemical corrosion \cite{Jiang1999}. Its potential in high-yield production of diverse devices paves the way for advancements in sensing \cite{halg2021membrane} and quantum technology \cite{Castelletto2020}.

In this work, we demonstrate wafer-scale amorphous films that harbor an ultimate tensile strength over 10 GPa after nanostructuring, a regime that is conventionally reserved for ultra-strong crystalline and 2D materials. Using delicate nanofabrication techniques, we produce several different nanomechanical resonators that can accurately determine the mechanical properties of SiC thin films such as density, Young's modulus, Poisson ratio, and ultimate tensile strength. Notably, our highest measured tensile strength ($>$10 GPa) is comparable to the values shown for c-SiC \cite{Kwon2015} and approaching the experimental values obtained for double-clamped graphene nano-ribbon \cite{Goldsche2018}. We achieve mechanical quality factor up to $2\times 10^8$ with a-SiC mechanical resonators, and measure loss-tangents on par with other materials used in high-precision sensors. Beyond sensing, these strong films open up new possibilities in high-performance nanotechnology, including thin solar cell technologies \cite{Kohler2021}, mechanical sensing \cite{Sementilli2022}, biological technologies\cite{nguyen2017nanopore} and even lightsail space exploration \cite{Atwater2018}.

\section{Fabrication of amorphous SiC resonators}

In pursuit of thin film materials for nanomechanical resonators with low mechanical dissipation, high film quality and high tensile stress are desirable. The Low Pressure Chemical Vapor Deposition (LPCVD) technique is preferred for these requirements, since its low pressure and high temperature deposition environment ensures lower defect density and higher thermal stress. The non-stoichiometric LPCVD a-SiC films used in this paper are deposited with different gas flow ratios (GFR) between SiH$_2$Cl$_2$ and 5\% C$_2$H$_2$ in H$_2$ (GFR=2,3,4), various deposition pressures (170 and 600 mTorr), and on both silicon and fused silica substrates (\textbf{Table \ref{table:1}}). This variation of deposition parameters allows us to systematically characterize the mechanical properties of LPCVD a-SiC thin films. All a-SiC thin films were deposited for the same period of time (3 hours 20 minutes) at a temperature of 760$^{\circ}$C in order to better determine the effect of various deposition environments while maintaining the films in the amorphous form \cite{MoranaPhD2015}. With the fabrication process demonstrated in the Experimental Section, nanomechanical resonators made of a-SiC can be suspended over the substrates with high yield using dry etching processes due to their extremely high chemical selectivity.

\begin{figure}[t]
\centering
\includegraphics[width=0.5\textwidth]{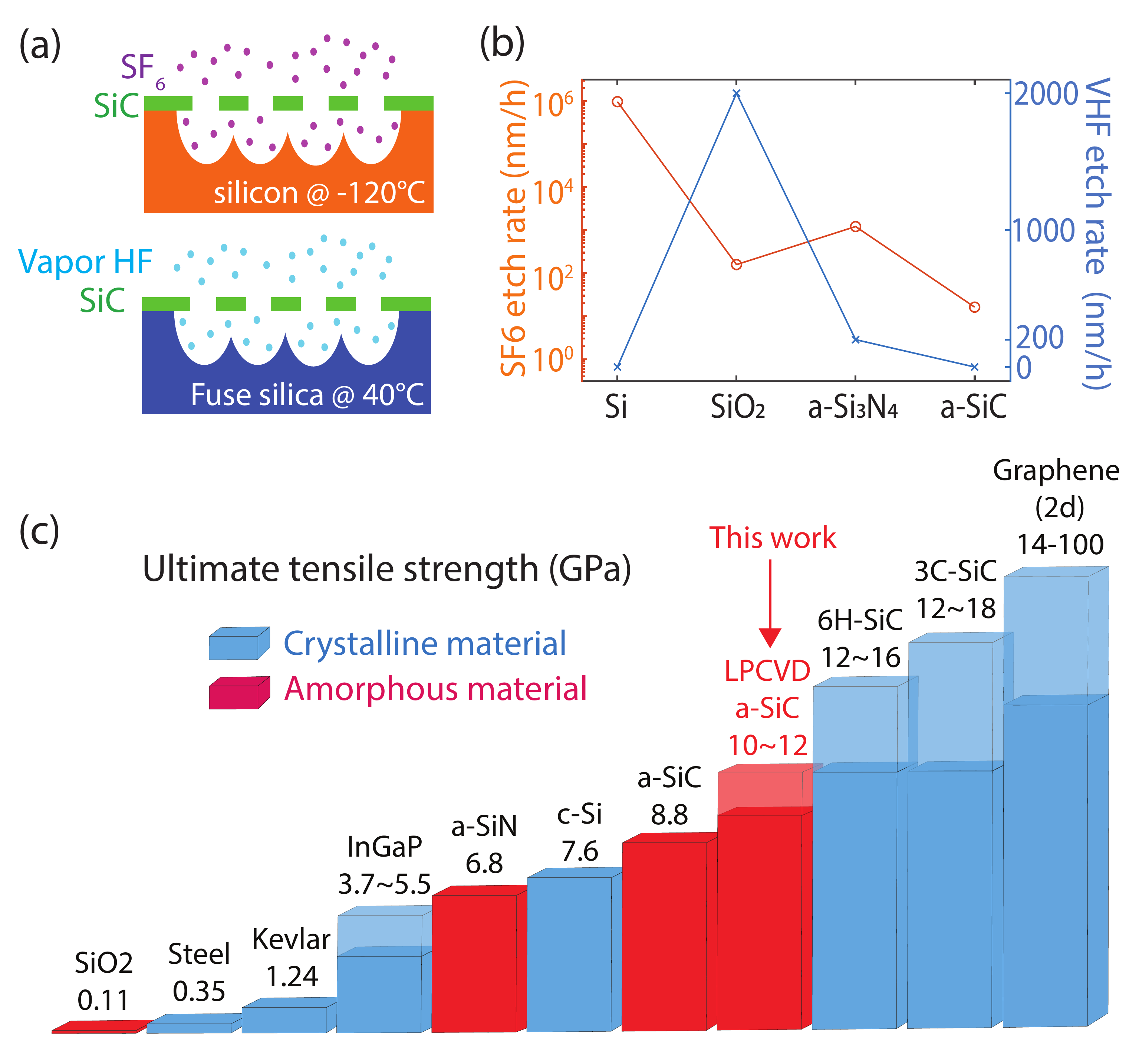}
\caption{(a) Schematic of dry under-cut processes with cryogenic SF$_6$ plasma isotropic etching (top) and vapor hydrofluoric acid etching (bottom), to suspend a-SiC nanomechanical resonators (green) on silicon (orange) and fused silica (blue) substrates, respectively. (b) Etch rates of cryogenic SF$_6$ plasma isotropic etching (log scale) and vapor hydrofluoric acid etching (linear scale), on four commonly used Si-based materials: Si, SiO$_2$, a-Si$_3$N$_4$, and a-SiC. (c) Ultimate tensile strength comparison between LPCVD a-SiC, crystalline (blue), and amorphous (red) materials. Reference: SiO$_2$ \cite{Chu2009}, steel \cite{Imran2018}, Kevlar \cite{Zhou2004}, InGaP \cite{Manjeshwar2022}, a-Si$_3$N$_4$ \cite{Norte2016,Bereyhi2019}, c-Si \cite{Sementilli2022, Chen2004}, a-SiC \cite{Cui2019}, 6H-SiC (crystalline) \cite{Kwon2015}, 3C-SiC (crystalline) \cite{Kwon2015}, graphene \cite{Goldsche2018,Rasool2013}. The solid and transparent colors of bars represent the lower and upper limits of the materials' ultimate tensile strengths, respectively.}
\label{fig1}
\end{figure}

Higher chemical stability and inertness of thin films is particularly useful for suspending nanostructures; usually a crucial and difficult step in fabricating free-standing structures.  LPCVD a-SiC films can be deposited on various substrates, patterned, and then suspended as nanomechanical resonators by removing the substrate underneath (i.e. undercutting). A high selectivity between the thin film and the substrate allows for higher yield and accuracy in fabricating suspended nanostructures. Similar to their crystalline counterpart, LPCVD a-SiC thin films have been reported to have very high chemical inertness to various wet etchants \cite{Morana2012}. Likewise, we found that a-SiC also has high chemical inertness to the widely used dry etchants, such as SF$_6$ isotropic plasma etching for silicon substrate, and vapor hydrofluoric acid etching for silicon oxide substrate, as illustrated in \textbf{Figure \ref{fig1}(a)}. The excellent chemical stability implies high selectivity between a-SiC and various commonly used substrates during undercutting, as shown in \textbf{Figure \ref{fig1}(b)}. Dry etchants are preferred for suspending high-aspect-ratio nanomechanical structures, since they help to avoid stiction during liquid etchant evaporation and thus improve the yield rate of working devices. To demonstrate the range of devices possible with higher chemical inertness, we fabricated nanomechanical resonators with continuous films down to 5nm, as shown in SEM pictures in Supporting Information \textbf{(F)}. In the Supporting Information \textbf{(H)} we show the measured chemical composition of these films, and demonstrate the chemical integrity of the 5nm SiC after etching.

The undercut method based on dry etchants introduces little perturbation to the suspended nano-structure, making it possible to perform delicate, on-chip tensile testing. The precision of the undercut method is pivotal for the reliability of material strength testing as it mitigates the risks associated with introducing fractures and defects during the loading/gluing of a material into a tensile testing setup, which could compromise the integrity of maximum tensile strength measurements. This chemical inertness of a-SiC serves a functional role in the fabrication process, allowing for a high degree of compatibility with various undercuts and ensuring the integrity of the nanostructures during the suspension process. The inertness contributes to the overall reliability and accuracy of the UTS measurements by inducing minimal forces during both fabrication and testing. In \textbf{Section \ref{sec4}}, we fabricate suspended nano-structures with different maximum tensile stresses to accurately determine the ultimate tensile strength (UTS) of a-SiC films. As a result, we demonstrate that a-SiC films have UTS up to 10-12 GPa, which are the highest among amorphous materials after patterning and are approaching the UTS of strong materials like c-SiC \cite{Kwon2015} and graphene nano-ribbons \cite{Goldsche2018}, both of which are known for their high UTS. The comparison of UTS between LPCVD a-SiC and other materials commonly used for nanomechanics is shown in \textbf{Figure \ref{fig1}(c)}.

\section{Mechanical property characterization with resonance method}
\label{sec3}

In order to design desired nanomechanical resonators with a specific thin film material, it is necessary to accurately characterize the material's mechanical parameters, such as film stress, Young's modulus, Poisson ratio and density. Various methods are developed to measure these parameters, including static methods, like nano-indentation \cite{Shuman2007, Kim2003} and dynamic methods like resonance response \cite{Klass2022, Barboni2018, Chirikov2020, Chen2000}. Many studies aiming to design high-performance nanomechanical resonators have relied on mechanical parameter values obtained from literature without considering potential variations of thin film properties due to different deposition environments, such as commonly used materials like a-Si$_3$N$_4$ \cite{Ghadimi2018, Shin2021}, c-Si \cite{Beccari2022}, and c-SiC \cite{Romero2020, Klas2022}. While these adaptations are usually reasonable and align well with experimental results, characterizing the exact parameters of the materials used would be beneficial when exploring the optimal performance of nanomechanical resonators \cite{Villanueva2014,Klass2022}. In this section, we present a simple and universal method to systematically characterize the important mechanical parameters of LPCVD a-SiC thin films.

The characterization flow of the method begins with measuring the thickness of the a-SiC thin film ($t$) after LPCVD deposition using a spectroscopic ellipsometer, which is an optical technique to confirm the thin film thickness and investigate its dielectric properties simultaneously. We then identify the film stress ($\sigma$) using the wafer bending method. After dicing the wafer into small chips, we pattern the a-SiC thin film and suspend it in the form of membranes, cantilevers, and strings with different lengths ($L$). The suspended nanomechanical resonators are measured with a laser Doppler vibrometer (LDV) in the vacuum environment down to $10^{-7}$ mbar. The measured resonant frequencies of the fundamental modes of the membranes ($f_{mem}$), cantilevers ($f_{can}$), and strings ($f_{str}$) can be fitted with their corresponding analytical expressions, which reveal the Young's modulus ($E$), Poisson ratio ($\nu$), and density ($\rho$) of the a-SiC thin film, respectively. During the fitting process, finite element method (FEM) simulation is used to describe the patterned resonators more precisely by taking into account the holes on the membranes and the overhangs from under-cutting adjacent to the cantilevers and strings. More detailed information about the measurements, analytical fitting, and simulations are shown in Supporting Information \textbf{(A)} and \textbf{(B)}.

\begin{figure*}[t]
\centering
\includegraphics[width=\linewidth]{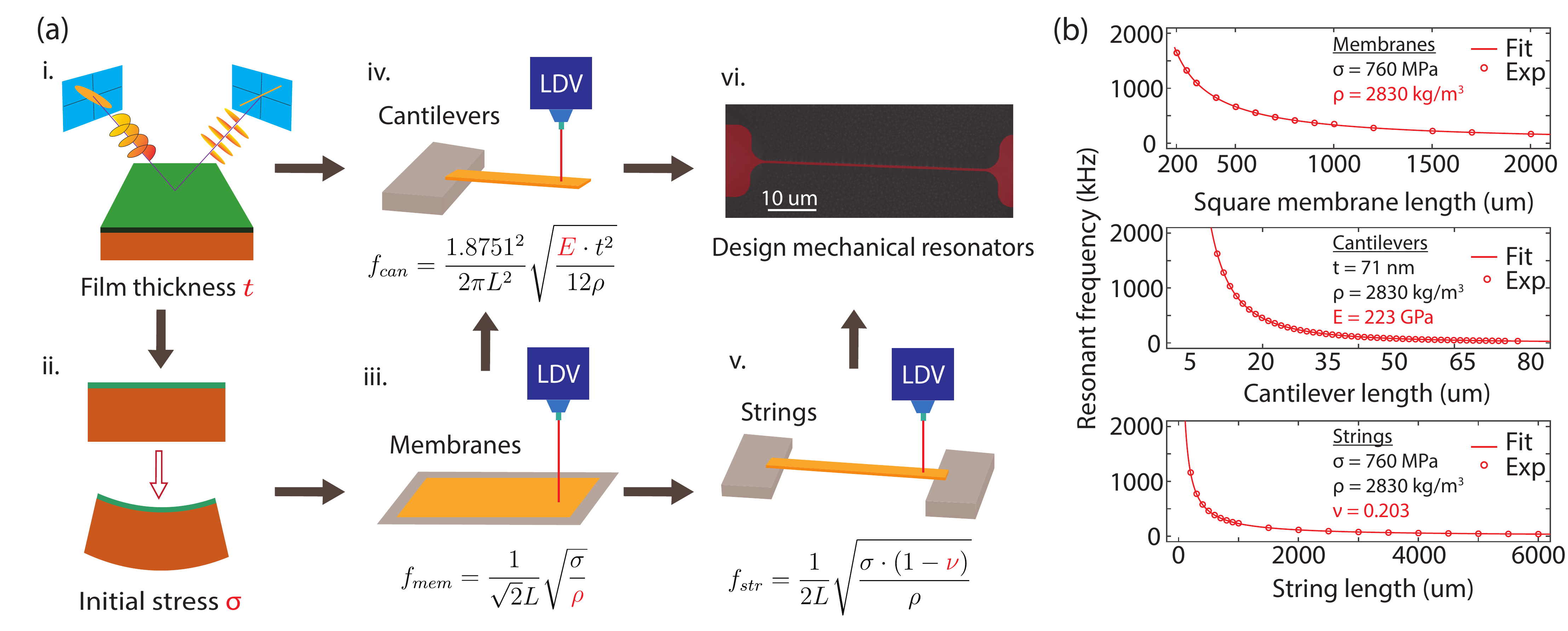}
\caption{ (a) Schematic of systematically characterizing mechanical properties of a tensile stress thin film material. (a-i): Measuring film thickness with ellipsometry. (a-ii): Measuring the film stress via wafer bending technique after film deposition. (a-iii) Extracting material density / (a-iv) Young's modulus / (a-v) Poisson ratio of a-SiC thin films by fitting resonant frequencies of square membranes / cantilevers / strings of different sizes, respectively. (a-vi): Designing a-SiC nano-mechanical resonators with desired performance. (b) Characterizing the mechanical properties of LPCVD a-SiC thin film (a-SiCR2) by numerically fitting the measured resonant frequencies of suspended resonators with different geometries and dimensions, including squared membranes (top), cantilevers (middle), strings (bottom). The resonant frequencies of the resonators mentioned above are measured with Laser Doppler Vibrometer (LDV, Polytec PSV-4).}
\label{fig2}
\end{figure*}

Using the above measurements, we can characterize the important mechanical parameters of LPCVD a-SiC thin films, then design and fabricate nanomechanical resonators with desired performance. Note that this straightforward and non-contact method can be universally applied to characterize the mechanical properties of other tensile thin film materials that can be fabricated into resonators with various geometries, i.e. cantilevers, strings, and membranes. This allows for quality control of thin films deposited in different batches or under varied deposition environments. As a result, nanomechanical resonators manufactured for various applications can be characterized in an efficient and economical manner, resulting in higher reliability for both industrial and academic applications. With all the relevant mechanical parameters accurately measured, we can fabricate a series of suspended devices specifically designed for characterizing the ultimate tensile strengths of a-SiC thin films in the following section.

\begin{table*}[ht]
\centering
  \begin{tabular}{@{}lllllllllll@{}}
    \hline
Recipe  \ \ \ \ \ \  &  GFR \ \ & P (mTorr) \ & Substrate \ \ \ \  & $t$ (nm) \  & $\sigma$ (MPa) \  & $\rho$\  (kg/m$^3$) & $E$ (GPa) \ \ & $\nu$ \ \ \ \ \ \ \  & $Q_0$\ \ \ \ \ \ & UTS (GPa)\\
\hline
a-SiCR2 & 2 & 600 & Silicon & 71 & 760 & 2830 & 223 & 0.203 & 5175 & 12.04 ($\pm$0.72)\\
a-SiCR2FS & 2 & 600 & Fused Silica & 82 & 1404 & 2555 & 187 & 0.222 & 4416 & - \\
a-SiC170 & 2 & 170 & Silicon & 86 & 636 & 2966 & 220 & 0.218 & 4485 & 10.27 ($\pm$0.62)\\
a-SiCR3 & 3 & 600 & Silicon & 81 & 960 & 2962 & 210 & 0.199 & 4692 & 11.12 ($\pm$0.45)\\
a-SiCR4 & 4 & 600 & Silicon & 137 & 670 & 3087 & 200 & 0.162 & 1035 & $<$3.5\\
    \hline
  \end{tabular}
  \caption{LPCVD a-SiC deposition parameters and the corresponding mechanical properties. From left to right columns, recipe name, gas flow ratio (GFR), deposition pressure P, thickness $t$, deposition stress $\sigma$, density $\rho$, Young's modulus $E$, Poisson ratio $\nu$, intrinsic quality factor $Q_0$ (per 100nm), and ultimate tensile strength (UTS), are shown. All a-SiC recipes have a deposition time 3 hours 20 mins, except a-SiCR2FS (3 hours 47 mins). Note that the mechanical properties of a-SiCR2FS are characterized with alternative methods presented in Supporting Information \textbf{(A5)}.}
\label{table:1}
\end{table*}

\section{Ultimate Tensile Strength of amorphous SiC}

\label{sec4}
\begin{figure*}[ht]
\centering
\includegraphics[width=\linewidth]{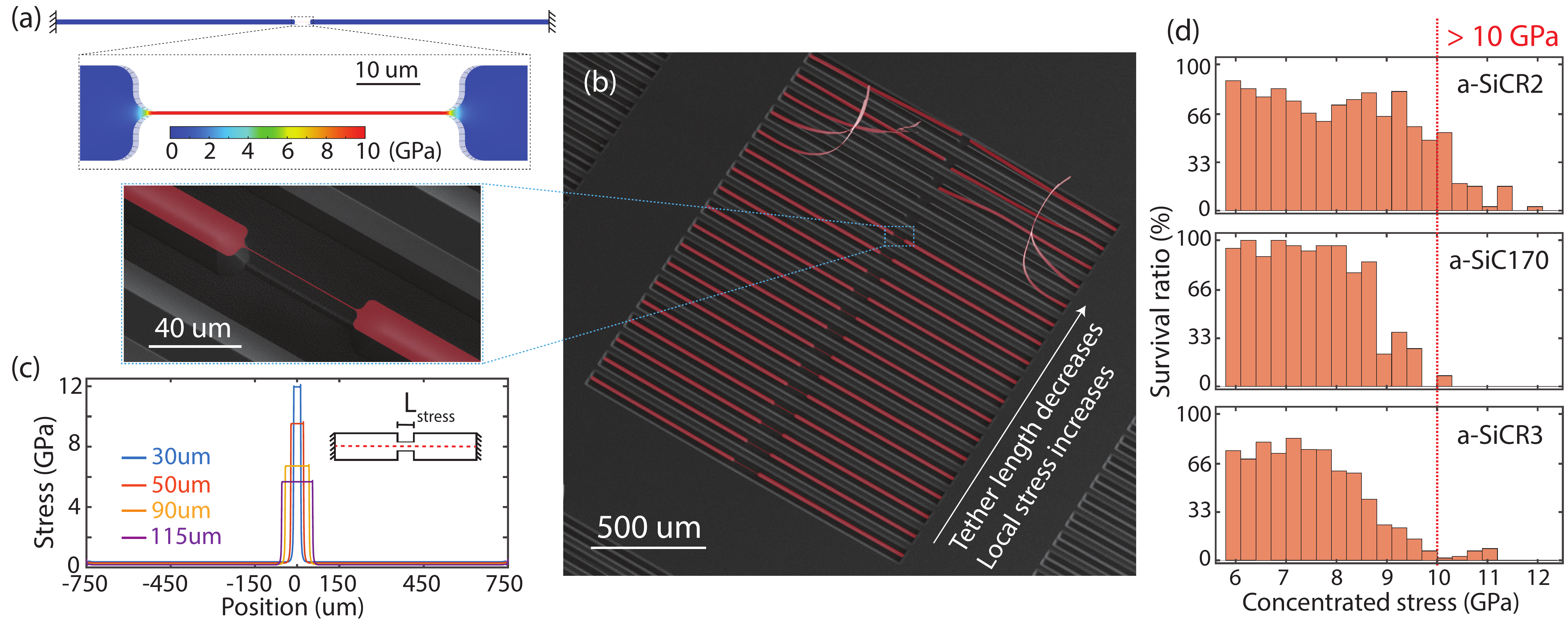}
\caption{Tensile test experiment to measure the ultimate tensile strength of a-SiC thin films (a) Simulated stress profile of a hourglass-shaped geometry of 50 um tether length made with a-SiCR2, the tensile stress is concentrated at the middle narrow tether up to 10 GPa, different maximum stress can be obtained with different tether lengths. (b) SEM image of a pad of a-SiCR2 hourglass-shaped structures with 18 different tether lengths, from 30 to 115 um. Below a certain tether length, the maximum stress surpass the ultimate tensile strength of the material, and the tethers break in the middle region after under-cutting, indicating the ultimate tensile strength of a-SiC. A zoom-in view of the hourglass-shaped geometry with a 50 um tether length is shown on the left. (c) The stress profiles along the hourglass-shaped geometries with different tether lengths. (d) Survival ratios of hourglass-shaped geometries with maximum stress correspond to different stress interval (orange columns) for a-SiCR2 (top), a-SiC170 (middle) and a-SiCR3 (bottom). The maximum stresses shown for unbroken devices fabricated with a-SiCR2/a-SiC170/a-SiCR3 are 12.04/10.27/11.12 GPa, respectively.}
\label{fig3}
\end{figure*}

Ultimate tensile strength (UTS), or sometimes coinciding with yield strength for brittle materials such as a-SiC \cite{Cui2019}, describes the maximum tensile stress a material can endure before breaking while being stretched. High UTS has been shown for nanowires fabricated with various materials, whose small cross-section areas minimize the appearance of defects \cite{Wang2017, banerjee2018ultralarge}, and for nanomechanical membranes without nano-patterning to avoid the presence of rough sidewalls \cite{Shafikov2021}. However, both scenarios above do not allow for further shape modification, reducing interest in their potential for various applications. While the crystalline form of materials usually tend to be mechanically stronger than their amorphous forms due to long-range order, examples such as glassy metal \cite{Demetriou2011} and synthesized AM-III carbon \cite{Zhang2022} demonstrate extraordinary mechanical properties comparable to their celebrated crystalline counterparts in terms of fracture toughness and yield strength, or hardness and compressive strength, respectively. This correspondence remains between c-SiC and a-SiC. While c-SiC has shown a UTS as high as 12-18 GPa via micro-pillars so far \cite{Kwon2015}, a-SiC nanowires have been measured to have a UTS up to 8.8 GPa via a tensile test with its two ends fixed by silver epoxy \cite{Cui2019}, which is higher than the ones shown for LPCVD a-Si$_3$N$_4$ (6.8 GPa \cite{Bereyhi2019}) and Si (7.6 GPa \cite{Sementilli2022}).

With the aim of characterizing the design space of nanomechanical resonators using LPCVD a-SiC thin film, we characterized its UTS by geometrically tapering the suspended a-SiC thin film in order to concentrate the tensile stress up to the fracture point. Unlike other tensile test methods \cite{namazu2023mechanical}, the presented method allows to determine the UTS of the tensile nanostructured film accurately, while avoiding the ambiguity caused by external loads, glues, and limitations of nano-fabrication, e.g. limited accuracy of nano-patterning and stiction during wet undercut. With the mechanical parameters characterized in \textbf{Section \ref{sec3}}, a-SiC hourglass-shaped devices consisting of a short and narrow tether surrounded by long and wide pads on both sides which are designed and suspended to measure the UTS of LPCVD a-SiC thin films. The devices have a total length of 1500um, with pads on both sides that have a width of 15 um, and the middle tethers that have varying lengths and a width of 500 nm, as shown in \textbf{Figure \ref{fig3}(a)}. After being suspended with dry etchants, the tensile stress on the hourglass-shaped device will redistribute and result in an increase of stress on the middle tether due to the pulling of the pads caused by residual stresses arising from the fabrication. The re-distributed stress profile in \textbf{Figure \ref{fig3}(a)} is obtained via finite element method (FEM). The devices are designed to have varying tether lengths from short to long, which are then arranged adjacently as shown in \textbf{Figure \ref{fig3}(b)}.

To establish a force equilibrium between the tether and the pads on each device, the ratio between the tensile stresses on the tether and the pads is inversely proportional to the ratio between their widths, combined with a small proportion of the lengths between the two, which enhances the strain (percentage of elongation) on the tether, the tensile stress on the tether in our hourglass-shaped devices can be significantly amplified during stress relaxation after suspending. As shown with FEM simulation in \textbf{Figure \ref{fig3}(c)}, devices with shorter tether lengths contain higher maximum concentrated tensile stresses on the tethers. This method allows the determination of the UTS of the nanostructured a-SiC thin films by counting the number of surviving devices after suspension. As shown in \textbf{Figure \ref{fig3}(b)}, a series of hourglass-shaped devices are fabricated with a-SiCR2. The 18 devices have tether lengths ranging from 30 to 115 um, corresponding to stresses from 12.53 to 5.97 GPa, respectively. The adjacent devices have tether lengths that differ for 5 um, the shorter the tethers are, the larger difference in concentrated stress the devices contain, e.g. the concentrated stress difference between devices with 115 um and 110 um tether lengths is 0.18 GPa, while one between devices with 35 um and 30 um is 0.72 GPa. In the case of each a-SiC thin film, the survival rate of each tensile interval shown in \textbf{Figure \ref{fig3}(d)} is determined, by employing 36 to 72 devices for testing.

The survival of the suspending hour-glass-shaped device with the tether length below 50 um, corresponds to a UTS above 10 GPa for a-SiCR2. Similarly, we can identify the UTS for all a-SiC thin films used in this study to be higher than 10 GPa, as shown in the histograms of ratios of survival devices in \textbf{Figure \ref{fig3}(d)}. The histograms also show that, with relatively higher deposition pressure and lower gas flow ratios, a maximum UTS up to 12 GPa can be achieved with a-SiCR2, which is almost twice that of the UTS shown for nanostructured LPCVD a-Si$_3$N$_4$ films. The measured UTS of a-SiCR4 is below 3.5 GPa, which is not attractive for further characterization. In the future, with a larger number of fabricated devices and a denser range of tether lengths, one can determine the UTS of the LPCVD a-SiC thin films more precisely. In practice, the nanopatterning with electron beam lithography can readily achieve an accuracy of 10 nm, which allows for the method's accuracy to be as low as 1.2 MPa on a-SiCR2, i.e. an error of less than 0.2\% when measuring the UTS. Higher UTS is found for recipes deposited with lower gas flow ratios (a-SiCR2/3/4), which might due to a higher carbon composition in the thin film \cite{MoranaPhD2015}, and C-C chemical bonds are stronger than Si-C and Si-Si bonds \cite{grunenberg2001intrinsic}. For a-SiC films deposited with different pressure, a-SiCR2 (600 mTorr) is found to have a higher UTS, while a-SiC170 (170 mTorr) exhibits better yield under lower concentrated stresses as shown by the survival rates. According to the relationship between strength and Young's modulus $E$ of SiC shown in \cite{Cui2019}, UTS (or fracture strength) is 5.3\% of $E$, therefore the theoretically predicted UTS for a-SiCR2/a-SiC170/a-SiCR3, are 11.82/11.66/11.13 GPa, respectively, matching well with the experimentally extracted data from the survived devices 12.04/10.27/11.12 GPa shown in \textbf{Table \ref{table:1}}. The small offset for the values of a-SiC170 may be due to its rougher surface as shown in Supporting Information \textbf{(G)}.

With strain engineering techniques, one can amplify the mechanical quality factor $Q=D_Q \cdot Q_0$ of a nanomechanical resonator by boosting their dissipation dilution factor $D_Q$, where $Q_0$ is the intrinsic quality factor of the thin film material \cite{villanueva2014evidence,Ghadimi2018}. Since the upper bound for $D_Q$ of a nanomechanical string vibrating at a certain frequency $\omega$ is given by $D_Q \leq 12E \epsilon^2_{UTS}/(\rho t^2 \omega^2)$, where $\epsilon_{UTS}$ denotes the UTS of the thin film material \cite{Fedorov2019}, thin film materials with higher UTS and lower thickness are advantageous to obtain a higher $D_Q$. Among all a-SiC thin films shown in this work, a-SiCR2 is the most promising one to maximize the $Q$ factor, thanks to its high $Q_0$ and UTS. The superior chemical resistivity of a-SiC enables the fabrication of thin films into suspending resonators with a thickness as low as 5 nm (shown in Supporting Information \textbf{(F)}). This combined with its elevated ultimate tensile strength $\epsilon_{UTS}$, which measures above 10 GPa in thicker films, makes a-SiC string resonators highly promising in achieving a supreme upper bound for $D_Q$ at a certain frequency $\omega$.

\section{Intrinsic Quality Factor and High Q Mechanical Resonators}

\begin{figure*}[t]
\centering
\includegraphics[width=\linewidth]{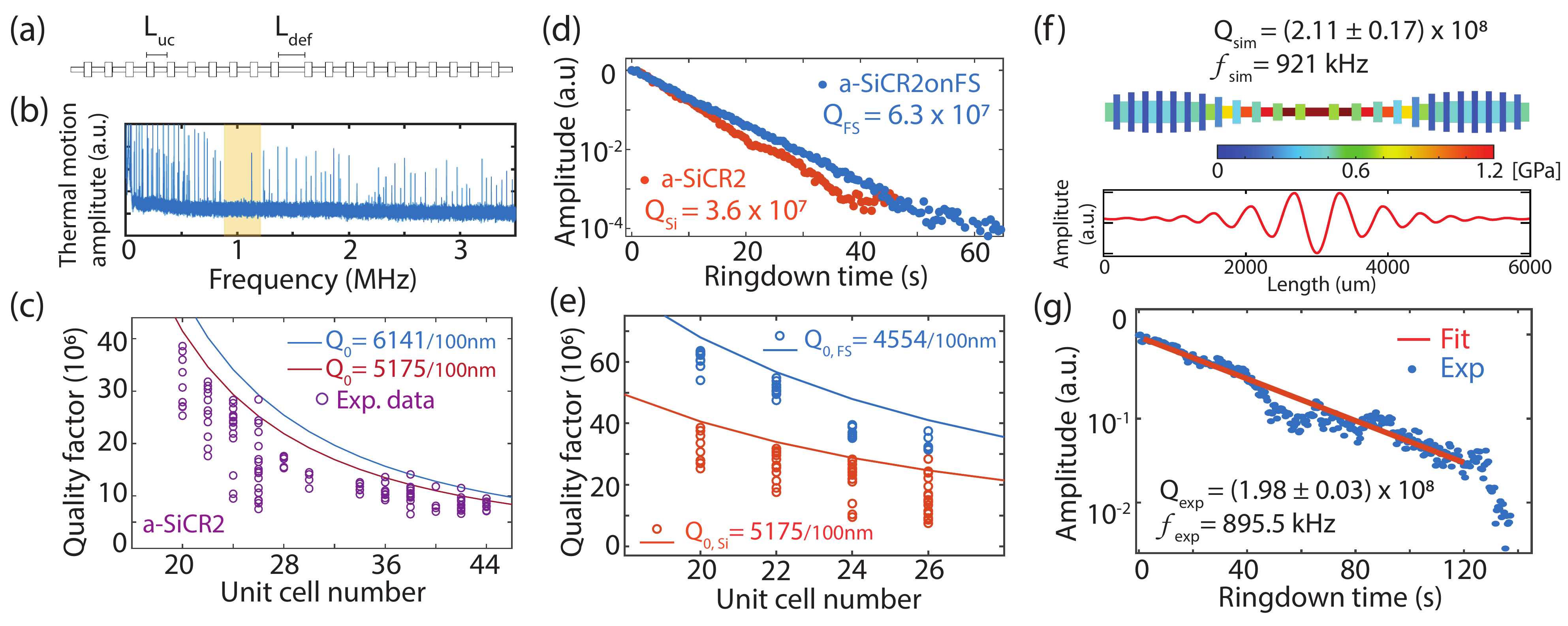}
\caption{Intrinsic quality factor characterization and high-Q factor a-SiC nanomechanical resonators optimized using Bayesian optimization. (a) The geometry of the defect mode, of a 20 unit-cell (UC) PnC nanostring with unit cell length L$_\mathrm{uc}$ and defect length L$_\mathrm{def}$. (b) The measured frequency spectrum of a 20 UC PnC nanostring made with a-SiCR2. The yellow shaded area represents the engineered phononic band gap. (c) The intrinsic quality factor $Q_0$ of a-SiCR2 is measured with 20 to 44 UC PnC nanostrings. (d) Ringdown measurements of 20 UC PnC nanostring made with a-SiCR2 (orange) and with a-SiCR2FS (blue). (e) Comparison of the intrinsic quality factors $Q_0$ of a-SiCR2 (orange) and a-SiCR2FS (blue) with PnC nanostrings of 20 to 26 UC. The hollow rings and the solid line represent the measured Q factors and the numerical fittings respectively. (f) The stress distribution (top) and mode shape of the defect mode (bottom) of the 6 mm tapered a-SiCR2 PnC nanostring optimized by Bayesian optimization. The optimized tapered PnC nanostring has 24 unit cells and a maximum stress of 1.2 GPa concentrated on the middle. (g) Ringdown measurement of the optimized tapered PnC nanostring, high-Q factor up to $Q_{exp}=1.98\times10^8$ is measured.}
\label{fig4}
\end{figure*}

In this section we characterize the intrinsic quality factor $Q_0$ of LPCVD a-SiC, then design and fabricate high-Q nanomechanical resonators with it. High mechanical quality (Q) factor nanomechanical resonators are desirable for various applications, ranging from precise force/acceleration sensing \cite{halg2021membrane,Krause2012}, microwave-to-optical conversion \cite{bagci2014optical}, to quantum optomechanics \cite{teufel2009nanomechanical,Guo2019}. Following the method introduced by LIGO \cite{Gonzalez1994}, the field of strain engineering is advancing rapidly, boosting the Q factor of nanomechanical resonators by several orders of magnitude. A variety of strategies have been proposed aiming to improve the Q factors of tensile-loaded nanomechanical resonators. These include patterning 2D geometries appropriately \cite{Norte2016,Fedorov2020,Bereyhi2022,Shin2021,Tsaturyan2017,Li2023}, modifying mass distribution \cite{Hoj2022,Pratt2021} and mode of interest (e.g., from fundamental to higher order or from flexural to torsional modes \cite{Pratt2021}), in-situ annealing for surface cleaning \cite{Sadeghi2021}, as well as cooling down to cryogenic temperatures \cite{Gisler2022,Beccari2022}. The methods mentioned above can benefit from utilizing the LPCVD a-SiC thin film we characterized in this work, due to its high deposition film tensile stress, superior chemical resistivity, and impressive ultimate tensile strength.

The intrinsic quality factors $Q_0$ of a-SiC thin films are identified by experimentally measuring the Q factors of phononic crystal (PnC) nanostrings \cite{Ghadimi2018}, whose many spurious loss mechanisms are eliminated, and dissipation dilution factor $D_Q$ is well defined, leading to an expected intrinsic Q factor $Q_0=Q/D_Q$. For thin nanomechanical resonators, $Q_0$ can be assumed to depend linearly on the film thickness, since it is predominantly determined by surface loss rather than bulk loss \cite{villanueva2014evidence}. We fabricate a series of uniformly corrugated high-aspect-ratio (PnC) nanostrings with a length of 4 mm, varying unit-cell lengths $L_{uc}$ and defect lengths $L_{def}$ in the middle (\textbf{Figure \ref{fig4}(a)}), leading to PnC nanostrings with unit-cell numbers from 20 to 44. The widths of the wide and narrow parts of the nanostrings are 3 um and 1 um respectively. The vibration amplitude of the nanostrings as a function of frequency is acquired (\textbf{Figure \ref{fig4}(b)}) with a custom balanced homodyne detection interferometer at the vacuum environment of $4\times 10^{-9}$ mbar (see schematic in Experimental Section). Using the ringdown method, the Q factors of defect modes for each PnC nanostring are measured. For example, the ones of 10 unit-cells PnC nanostrings fabricated wtih a-SiCR2 and a-SiCR2FS are plotted in \textbf{Figure \ref{fig4}(d)}. Using FEM simulation, the dilution factor $D_Q$ of each PnC nanostring geometry can be numerically calculated. Together with the Q factors of the corresponding nanostring measured experimentally, the intrinsic Q factor $Q_0$ of the different a-SiC thin films are determined. For example, the $Q_0$ of a-SiCR2 and a-SiCR2FS are shown in \textbf{Figure \ref{fig4}(c)} and \textbf{(e)} respectively. The $Q_0$ of the other a-SiC films are shown in \textbf{Table \ref{table:1}}, and the corresponding measurement data can be found in Supporting Information \textbf{(D)}. In order to compare the $Q_0$ of a-SiC thin films with different thicknesses, we present them with the unit $Q_0$ per 100 nm, as $Q_0$ of a thin film is shown to be a function of thickness \cite{Villanueva2014}. Deposited with the same recipe, a-SiCR2 (5175/100 nm) and a-SiCR2FS (4554/100 nm) have similar $Q_0$. The similar performance on the transparent substrate allows for integrating high-Q nanomechanical sensors into free-space optical systems in a practical manner. By reasonably assuming the films have similar mechanical properties on different substrates, a-SiCR2FS is measured to have a deposition stress of 1596 MPa, a factor of two higher than a-SiCR2 due to a larger thermal expansion coefficients difference between the a-SiC thin film and fused silica substrate. Worth noting is that the $Q_0$ of a-SiCR2 is the highest among all LPCVD a-SiC investigated, indicating that a lower gas flow ratio (GFR=2), i.e., more carbon content \cite{MoranaPhD2015}, and a moderate deposition pressure (600 mTorr) is beneficial to have better film quality.

To exploit the sensing potential of LPCVD a-SiC, we designed and optimized a tapered PnC nanostring with a length of 6mm and a thickness of 71 nm using a a-SiCR2 thin film. Bayesian optimization \cite{Shin2021} was used to find designs with high Q-factor -- more details can be found in Supporting Information \textbf{(E)}. This simulation-based optimization is largely possible due to the accurate characterization of the material properties of the a-SiC thin films in previous sections. As shown in \textbf{Figure \ref{fig4}(f)}, the optimized PnC nanostring consists of fixed 24 unit cells with different widths and lengths, leading to a stress concentration of up to 1.2 GPa towards its center part. Within the phononic bandgap generated by the optimized tapered PnC nanostring, a soft clamped defect mode with a simulated Q factor $Q_{sim}=2.11\pm 0.17\times 10^8$ appears at the frequency of $f_{sim}=921$ kHz, as shown at the bottom of \textbf{Figure \ref{fig4}(f)}. The optimized tapered PnC nanostring was fabricated based on the design at the top of \textbf{Figure \ref{fig4}(f)}, and it was measured at an interferometer under ultra-high vacuum of $4\times 10^{-9}$ mbar. As a result, an high Q factor mechanical mode with $Q=(1.98\pm0.03)\times 10^8$ was measured experimentally at a frequency of $f=896$ kHz at room temperature, shown by its ringdown curve plotted in \textbf{Figure \ref{fig4}(g)}. This result demonstrates, for the first time, a mechanical quality factor exceeding $10^8$ for silicon carbide nanomechanical resonators, as predicted by simulation. This also suggests that future design strategies to enhance resonator performance can be carried out using the LPCVD a-SiC thin films.

In addition, the quality factor-frequency product of the optimized LPCVD a-SiC tapered PnC nanostring is $Q\times f = 1.791 \times 10^{14}$, which is significantly higher than the quantum limit $Q\times f = 2\pi k_B T /\hbar = 6.24\times 10^{12}$. This paves the way towards engineering quantum states in room temperature environments \cite{Guo2019,Norte2016}. This high quality factor of the nanomechanical resonator with an effective mass $m_{eff}=1.27\times 10^{-13}$ kg corresponds to a force sensitivity of $\sqrt{S_{F}}= \sqrt{4k_B T m_{eff} \cdot 2\pi f/Q } =$ 7.7 aN/Hz$^{1/2}$ at room temperature, which is comparable to a typical atomic force microscope cantilever operating at liquid helium temperature. With the high quality factor shown above, LPCVD a-SiC is shown to be the third material that can reach $Q>10^8$ at room temperature using strain engineering, after conventional a-Si$_3$N$_4$ \cite{Ghadimi2018} and strained silicon \cite{Beccari2022}. Moreover, the superior chemical and mechanical properties of LPCVD a-SiC allow for the fabrication of thinner and stronger resonators, enabling it to be more compatible with the dissipation dilution method. With advantages such as a relatively simple and low-cost fabrication process, compatibility with various substrates, including transparent ones, its high-Q performance, LPCVD a-SiC is a promising material for fabricating commercial mechanical sensors.

\section{Conclusion and outlook}

In summary, our study has uncovered an amorphous silicon carbide thin film with a ultimate tensile strength above 10 GPa, the highest value ever measured for a nanostructured amorphous material and approaching the experimental values shown by graphene nano-ribbons \cite{Goldsche2018}. Their robustness to chemicals allow us to fabricate nanostructures with very high fidelity even when their geometries make them delicate high-aspect-ratio structures. This ability to reliably produce structures also allow us to measure the film's mechanical properties with high precision. We deposit amorphous silicon carbide in varying deposition conditions and substrates to understand new approaches towards increasing ultimate yield strength. Then using the a-SiC with the highest UTS, we designed and fabricated a variety of well-understood nanostructures such as cantilevers, membranes and doubly-clamped strings to measure the thin films mechanical properties such as density, Young's modulus, Poisson ratio, and mechanical loss tangent. For the latter we employ nanostrings patterned with phononic bandstructures which conventionally show some of the lowest mechanical dissipations in literature, and this allows us to measure very low mechanical dissipation. The a-SiC nanostrings support soft-clamped mechanical modes with quality factors exceeding $10^8$ at room temperature; a new regime for SiC devices and on par with the state-of-the-art SiN resonators. This corresponds to a high force sensitivity of $\sqrt{S_{F}}=$ 7.7 aN/Hz$^{1/2}$. We demonstrate a robust characterization process based on the simple fabrication and optical techniques which does not rely on complex tension loading setups. 

From molecular analysis of chemical composition (shown in the Supporting Information \textbf{(H)}), it appears that the elevated tensile strength in our a-SiC films is likely attributed to a higher prevalence of robust C-C bonds compared to weaker Si-Si bonds. This insight not only sheds light on the intricate interplay of bond structures and deposition parameters in determining mechanical resilience but also lays an understanding for further exploration in harnessing the unique properties of amorphous materials.

The discovery of this amorphous SiC material represents an advancement in the field of high-strength material science which is conventionally dominated by crystalline and 2D materials. However, our findings demonstrate that amorphous materials have the potential to surpass crystalline materials in certain applications due to their inherently isotropic mechanical properties, which allow for more design freedom and ease of fabrication. The high ultimate tensile strength of this amorphous material is particularly attractive for mechanical sensors, as it enables greater flexibility in strain engineering. This discovery opens up new possibilities for the use of amorphous materials in a variety of high-performance applications.

\begin{acknowledgments}
We wish to acknowledge Peter G. Steeneken, Gerard Verbiest, and Martin Lee for their helpful suggestions on the manuscript and their support of our project. We want to thank Satadal Dutta, Ali Sarafraz, Matthijs de Jong, Hanqing Liu for the helpful discussions. M.X. and R.N. also thank the staffs of both the Kavli Nanolab Delft and the Else Kooi Lab, in particular from Charles de Boer, for supporting our fabrication efforts, and from Hozanna Miro, for helping to perform the XRD measurement. The authours appreciate the help from P.R. Anusuyadevi and P. Gonugunta on the XPS measurement. This publication is part of the project, Probing the physics of exotic superconductors with microchip Casimir experiments (740.018.020) of the research programme NWO Start-up which is partly financed by the Dutch Research Council (NWO).This work has received funding from the EMPIR programme co-financed by the Participating States and from the European Union’s Horizon 2020 research and innovation programme (No. 17FUN05 PhotoQuant). R.N. would like to acknowledge support from the Limitless Space Institute’s I$^2$ Grant. Funded/Co-funded by the European Union (ERC, EARS, 101042855). Views and opinions expressed are however those of the author(s) only and do not necessarily reflect those of the European Union or the European Research Council. Neither the European Union nor the granting authority can be held responsible for them. 
\end{acknowledgments}

\appendix

\nocite{*}


\providecommand{\noopsort}[1]{}\providecommand{\singleletter}[1]{#1}%

\newpage
\ 
\newpage

\begin{widetext}

\begin{center}
\large{\textbf{SUPPLEMENTARY INFORMATION}}
\end{center}

\renewcommand{\theequation}{S.\arabic{equation}}
\renewcommand{\thesection}{}
\renewcommand{\thesubsection}{}
\renewcommand{\thetable}{S\arabic{table}}  
\renewcommand{\thefigure}{S\arabic{figure}}

\section*{Supporting Information \textbf{(A)}: Mechanical properties characterization of LPCVD a-SiC thin films using resonance method}

\begin{figure*}[h!]
\centering
\includegraphics[width=1\textwidth]{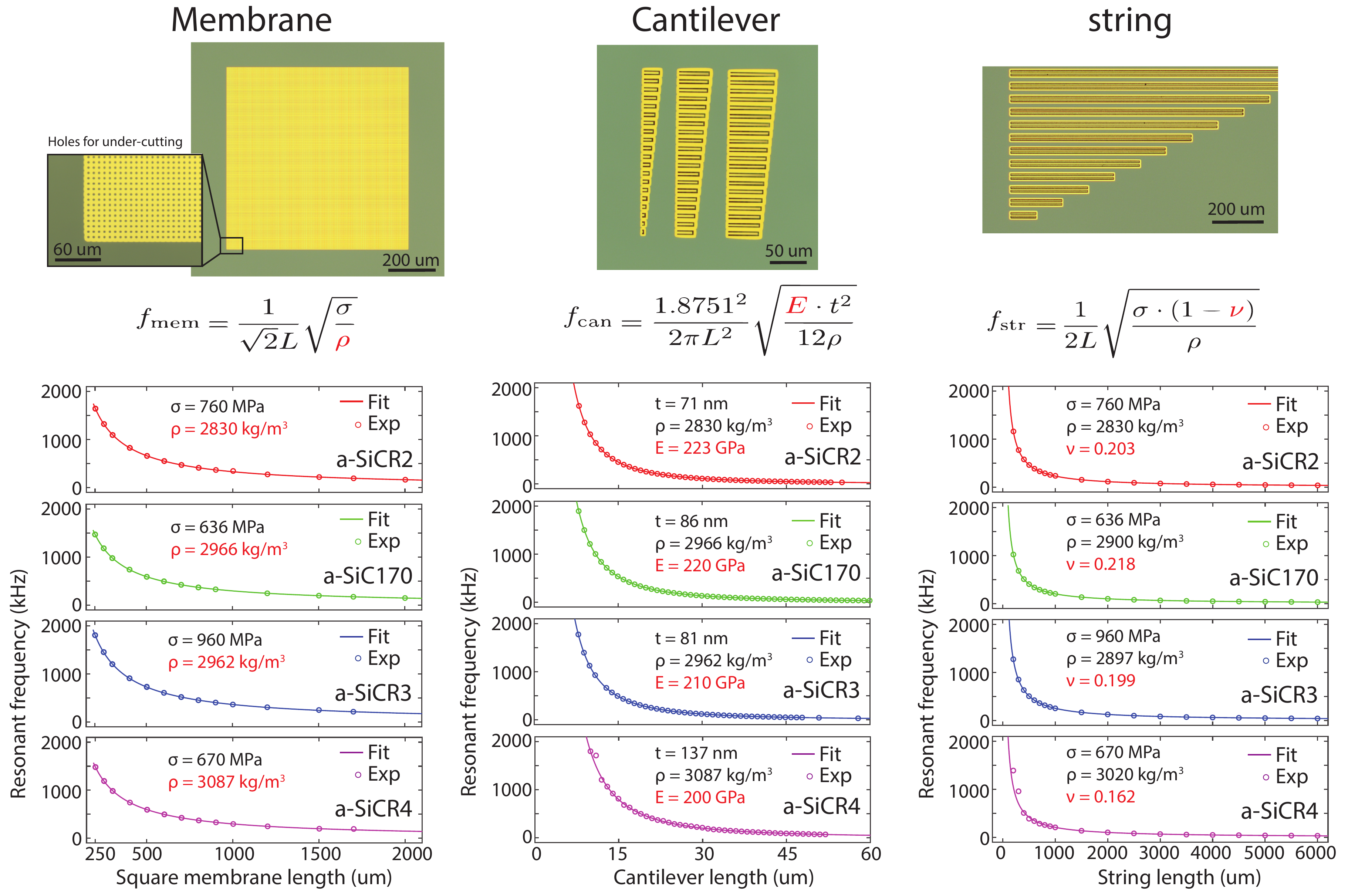}
\caption{Mechanical properties characterizations. After identifying the film stress and thickness with wafer bending and ellipsometry methods, the density $\rho$ / Young's modulus $E$ / Poisson ratio $\nu$ of a-SiC films deposited at different conditions are systematically measured with the resonant frequency of the a-SiC membranes (left column)/cantilevers (middle column)/strings (right column) made of the specific a-SiC film, as shown with images on the top. From top- to bottom of each row, resonant frequency data (hollow dots) and analytically fitting (solid line) for a-SiCR2/a-SiC170/a-SiCR3/a-SiCR4/a-Si$_3$N$_4$ thin films are shown respectively. The bottom row are measurement and fitting for the widely used material stoichiometric a-Si$_3$N$_4$ for reference purpose, whose fitted mechanical properties are close to the ones reported \cite{Shin2021,Ghadimi2018}, validating the generality of the method.}
\label{SfigMechPro}
\end{figure*}

\subsection*{A1. Characterization formulas for different mechanical resonators}

The characterization flow of the method start with measuring the a-SiC thin film thickness $t$ after the LPCVD a-SiC deposition using the Spectroscopic Ellipsometer (Woollam M-2000F). Then we identify the film stress by measuring the radius curvature $R_1$ of the silicon wafer before the deposition with the stress meter (Flexus, Toho), and measuring again the curvature $R_2$ after the deposition with the a-SiC on the backside of the wafer removed with CHF3/Ar plasma anisotropic etching. The film stress $\sigma$ can be determined from wafer bending method by Stoney's equation 
\begin{equation}
    \sigma=\frac{E_{sub}D_{sub}^2}{6(1-\nu_{sub})t}(\frac{1}{R_1}-\frac{1}{R_2}),
\end{equation}
where $E_{sub}$, $\nu_{sub}$ and $D_{sub}$ is the real component of Young's modulus, Poisson ratio and thickness of the substrate (silicon wafer in our case), respectively, and $t$ is the thickness of the a-SiC thin film. Apart from film stress, Young's modulus $E$, Poisson ratio $\nu$ and density $\rho$ of a-SiC are most relevant among all material properties to designing a-SiC resonator with targeted resonant frequency and stress distribution, which can be measured by patterning and then suspending the thin film as squared membrane, cantilevers and strings of different lengths $L$. After suspended, the nanomechanical resonators are measured with Laser Doppler Vibrometer (LDV, Polytec PSV-400), while they are placed in a vacuum chamber pumped down to 10$^{-7}$ mbar vacuum environment. After measuring the resonant frequencies of the squared membranes with lengths $L$ varying from 200 to 2000 um, we fit the measured data with the analytical formula for fundamental mode \cite{Wilson2009}
\begin{equation}
    f_{mem}\approx\frac{1}{\sqrt{2}L_{eff}}\sqrt{\frac{\sigma}{\rho_{eff}}},
    \label{membrane}
\end{equation}

where $L_{eff}=L+L_{oh}$ is the effective length includes the overhang size $L_{oh}$ generated during undercut, $\rho_{eff}=A_{corr}\times \rho$ is the effective density of the thin film, and $A_{corr}$ is the correction factor due to the arrays of holes on top for fast undercut, which in our case $A_{corr}=0.804$ calculated from COMSOL (corresponds to the holes with diameter 1.5 um are placed 3 um apart between adjacent centers, see \textbf{Figure \ref{SfigMechPro}}). We can therefore determine the density $\rho$ of the thin film since $\sigma$ and $L$ are known beforehand. The resonant frequencies of strings \cite{Buckle2021} with lengths from 200 to 6000 um is measured with LDV. The eigenfrequency of string resonators can be analytically formulated as
\begin{equation}
    f_{str,n} = \frac{n^2 \pi}{2L^2} \sqrt{\frac{E t^2}{12\rho}}\sqrt{1+\frac{12\sigma_{\text{1D}} L^2}{n^2 \pi^2 E t^2}}
\end{equation}

where $L$ is the length of the string, and needed to be modified into $L_{\text{eff}}$ due to the overhang from undercutting, $n$ is the eigenmode number, $\rho$ is the material density, $\sigma_{\text{1D}}=\sigma \times (1-\nu)$ is the tensile stress on the string, $\sigma$ is the film stress and $\nu$ is the Poisson ratio, $E$ is the Young's modulus and $t$ is the thickness of the film. In our case, a-SiC thin films have high tensile stress, which leads to $12\sigma_{\text{1D}} L^2 \gg n^2 \pi^2 E t^2$, and the formula of the fundamental mode is reduced to the form one can use to fit the measurement data
\begin{equation}
    f_{str}\approx\frac{1}{2L_{eff}}\sqrt{\frac{\sigma \cdot (1-\nu)}{\rho}},
\end{equation}
from which the Poisson ratio $\nu$ of a-SiC can be determined. Also the resonant frequency of cantilevers \cite{macho2015oscillations} with lengths from 7 to 80 um are also measured, and can be fitted to the analytical formula in the following form
\begin{equation}
    f_{can}\approx\frac{1.8751^2}{2\pi L_{eff}^2}\sqrt{\frac{E\cdot t^2}{12 \rho}},
\end{equation}
from which the Young's modulus $E$ of a-SiC can be determined.

\subsection*{A2. Aspect ratio requirement for membrane resonators}

Adapting from the analysis shown for plates under tension in the papers \cite{steeneken2021dynamics,castellanos2013single}, the total resonant frequency of squared membrane $f$ can be approximated as

\begin{equation}
    f \approx \sqrt{f^2_{membrane}+f^2_{plate}},
\end{equation}

here $f_{membrane}$ is the fundamental frequency of the square membrane-like resonator \cite{Wilson2009}, 
\begin{equation}
    f_{membrane} = \frac{1}{\sqrt{2}L}\sqrt{\frac{\sigma}{\rho}},
\end{equation}

and $f_{plate}$ is the fundamental frequency of the square plate-like resonator with clamped boundary condition is \cite{duvigneau2016vibration}

\begin{equation}
    f_{plate} = \frac{\pi\cdot t}{L^2}\sqrt{\frac{E}{12(1-\nu^2)\rho}},
\end{equation}

where $L$ is the length, $t$ is the thickness, $\sigma$ is the tensile stress, $\rho$ is the density, $E$ is the Young's modulus, and $\nu$ is the Poisson ratio of the square resonator.

To answer the reviewer's comment: Equation (S2) is valid with high accuracy, if the requirement
\begin{equation}
    f_{membrane}^2 \gg f_{plate}^2,
\end{equation}

is met. To make it make clear, this put a requirement on the aspect ratio of the membrane resonators:\newline

\begin{equation}
    \frac{L}{t}\gg \pi \sqrt{\frac{E}{6\sigma(1-\nu^2)}},
\end{equation}

In our case, using properties of a-SiC170 thin film as an example, the membranes fabricated with it should have dimensions that satisfy $L/t\gg 25$, which is true even with the smallest membranes we used in this work, $200 \mu m/86 nm=2326 \gg 25$. With the above argument, we show the use of equation (S2) is valid for our structures.\newline

\subsection*{A3. Methodology verification with LPCVD Si$_3$N$_4$ with various thicknesses}

To show the strength of the approach, we have redone the entire characterization for 5 different thicknesses to show that we can get results in line with what is known about a-SiN. We specifically deposited LPCVD stoichiometric silicon nitride to make sure the material composition was consistent between films, including the same furnace and deposition parameters. It should be noted that a-SiN thin-films' properties can vary between different furnaces, deposition recipes, testing methodologies, etc. To our knowledge, this is one of the more comprehensive measurements of Young's modulus, density, and Poisson ratio done for LPCVD stoichiometric Si$_3$N$_4$. These values tend to agree with values used in other rigorous studies of Si$_3$N$_4$ \cite{Fedorov2019} which uses values of $\sigma = 1.14~$GPa, $\rho =3100~$kg/m$^3$, $\nu = 0.23$, $E = 250~$GPa. We thank the reviewer for this suggestion as we feel it anchors the approach to a well-known material.


\begin{table}[ht]
\centering
  \begin{tabular}{@{}lllllllllll@{}}
    \hline
Name \ \ \ \ \   & Substrate \ \  & $t$ (nm) \ \ \   & $\sigma$ (MPa) \ \ \   & $\rho$ (kg/m$^3$) \ \ \   & $\nu$ \ \ \ \ \ \  \ \    & $E$ (GPa)\\
\hline
Si$_3$N$_4$-1 & Silicon & 50.04 & 1060 & 3159 & 0.255 & 259\\
Si$_3$N$_4$-2 & Silicon & 68.38 & 1080 & 3097 & 0.245 & 259\\
Si$_3$N$_4$-3  & Silicon & 89.56 & 1080 & 3072 & 0.232 & 261\\
Si$_3$N$_4$-4  & Silicon & 210.77 & 1080 & 3024 & 0.214 & 265\\
Si$_3$N$_4$-5  & Silicon & 339.30 & 1080 & 2981 & 0.218 & 264\\

    \hline
  \end{tabular}
  \caption{Mechanical properties characterization of stoichiometric Si$_3$N$_4$ thin films of different thicknesses, with the process flow shown in this work.}
\label{table:3}
\end{table}

\subsection*{A4. Thickness uniformity and mechanical property consistency of a-SiC film across the wafer}

To verify the film thickness and stress measurement on wafer-level is appropriate for the a-SiC thin films we studied in this work, we have conducted a number of additional experiments. Ellipsometry measurements were performed across the wafer to verify the uniformity of the a-SiC thin films’ thickness. Additionally, we assessed the mechanical properties of the a-SiC thin film using dies from various parts of the wafer as shown in Figure \ref{AMreply1}.

 Figure \ref{AMreply1} shows the uniformity in thickness across the wafer, substantiating the robustness of our measurement approach. The depicted color bar and measurement spots provide a detailed view of the a-SiCR2 thin film thickness at multiple points, illustrating a consistent thickness with a minor deviation of approximately 5nm over 100 mm.

Table \ref{table:2} outlines the mechanical properties of a-SiCR2 thin films, tested on three distinct chips located at varied positions on the wafer. The observed minimal variations in thickness, stress, density, Young's modulus, and Poisson's ratio across the different chips underscore the reliability and consistency of our wafer-scale assessments.

We hope these additional experiments and data addresses the valid concerns regarding the representativeness and validity of our measurements.

\begin{figure*}[ht]
\centering
\includegraphics[width=0.5\textwidth]{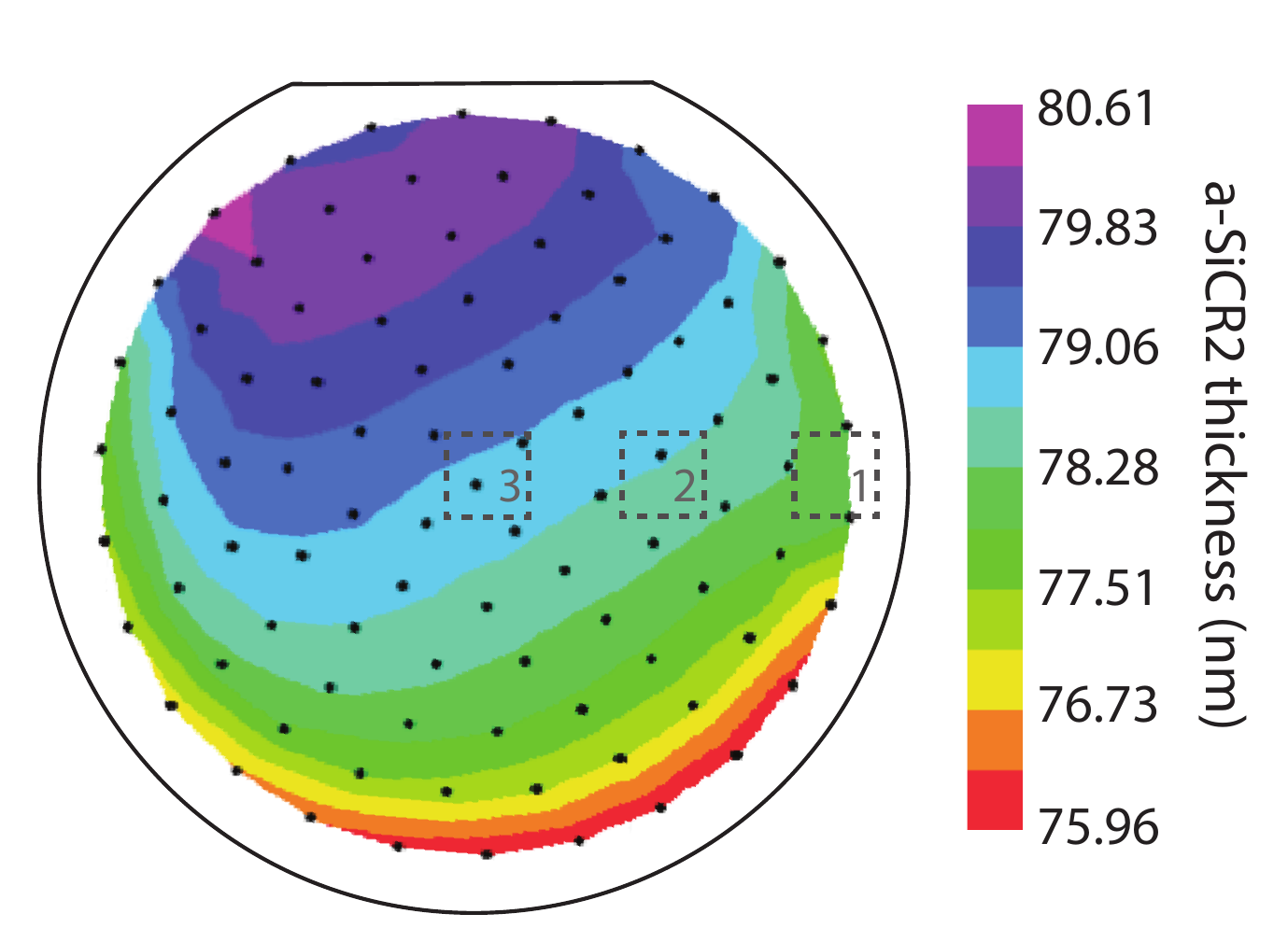}
\caption{Thickness distribution on the silicon wafer coated with 80nm a-SiCR2 thin film (deposition time: 3 hours 47 mins), measured with ellipsometry. The color bar represents the a-SiCR2 thin film thickness around the measurement spots (black dots). From the edge to the center of the wafer, locations of chip 1/2/3 of dimensions 10x10 mm$^2$ are shown, whose mechanical properties are characterized in Table \ref{table:2}.}
\label{AMreply1}
\end{figure*}

\begin{table}[ht]
\centering
  \begin{tabular}{@{}lllllllllll@{}}
    \hline
Location \ \ \ & Recipe \ \ \ \ \ \  & Substrate \ \ \  & $t$ (nm)\ \ \  & $\sigma$ (MPa)\ \ \  & $\rho$ (kg/m$^3$)\ \ \  & $\nu$ \ \ \ \ \ \ \ \ \   & $E$ (GPa)\\
\hline
Chip 1 & a-SiCR2 & Silicon & 78.47 & 760 & 2866 & 0.212 & 224\\
Chip 2 & a-SiCR2 & Silicon & 78.73 & 760 & 2857 & 0.212 & 225\\
Chip 3 & a-SiCR2 & Silicon & 79.04 & 760 & 2846 & 0.210 & 225\\

    \hline
  \end{tabular}
  \caption{Mechanical properties of a-SiCR2 thin films on three chips located differently across the wafer.}
\label{table:2}
\end{table}

\newpage
\subsection*{A5. Alternative method to measure mechanical properties of a-SiCR2 on fused silica substrate}

In order to check the mechanical properties of a-SiCR2 thin film deposited on fused silica substrate, which is transparent thus limits the validity of the method we presented in this work, we developed an alternative procedure to characterize the a-SiCR2 thin film on fused silica.

To measure the stress on the film, we firstly coated 40 nm metal (niobium titanium nitride) on top of the a-SiCR2 film on the front side of the double polished fused silica wafer to make the surface detectable for the stress meter, then we use directional plasma etching to remove the a-SiCR2 thin film on the back side of the wafer. After measuring and recording the bending radius before and after the etching, we diced the wafer into chips, and subsequently measure the thickness of the a-SiCR2 thin film on top with the scanning electron microscopy (SEM), which is found to be found to be 82$\pm$1 nm. By knowing the thickness of the film, we calculate the tensile stress of the a-SiCR2 film on the back side, and infer a similar value for the film on the front side, due to the uniformity of the LPCVD deposition. This process can be seen as depositing a compressive film of the same thickness and opposite stress on both sides of the wafer. We found the average stress of the a-SiCR2 film on fused silica to be 1404 MPa. Since the etch rate of vapor HF is hard to control, we notice that there is a large undercut on the margins of the membranes (13.6 um). For the sake of accuracy, we use FEM to simulate the real shapes of the membranes, and find that with the density 2555 kg/m$^2$, the membrane frequency can be fitted the best. The Poisson ratio of the film is then found to be 0.222 by fitting the frequency data of the strings, by applying the same method we used in the main text (Figure 2). Since cantilevers tend to collapse in our attempts to undercut with vapor HF, we resort to an alternative method to determine the Young's modulus of the film, namely using the frequencies of the torsional modes of the 1d PnC nanostring.

We firstly use the mechanical parameters we obtained above to fit the frequencies of the flexural modes up to mode number 20th. By comparing the measurement data and the FEM simulation data, we find a very good match between the two, with only $\Delta f_{flex}/f_{flex} = (f_{flex,data} - f_{flex, fit})/f_{flex,data}\times 100\%=0.16\%$ deviation found, as shown in Figure \ref{AMreply_SiCR2onFS}(c). Later on, we fit the Young's modulus $E$ of the material by changing it slightly in the FEM simulation, which allows us to obtain the $E$ that can fit the data of torsional mode frequency optimally, in our case $E=186$ GPa. Note that the frequency of the torsional modes instead of the flexural modes is used, because it is more sensitive to the change of $E$. As shown in \cite{Pratt2021}, the frequency of the $n$-th torsional mode of a uniform doubly-clamped string, with high aspect ratio $\{w,t\}\ll L$, is

\begin{equation}
    f_{tor,n} = \frac{n}{4\pi L}\sqrt{\frac{\sigma}{\rho}(1+\frac{4 Et^2}{\sigma\cdot w^2})},
\end{equation}

Here $w$ is the width of the string resonator. Unlike the frequency of the flexural mode, 
\begin{equation}
    f_{str,n} = \frac{n^2 \pi}{2L^2} \sqrt{\frac{E t^2}{12\rho}}\sqrt{1+\frac{12\sigma (1-\nu) L^2}{n^2 \pi^2 E t^2}},
\end{equation}
the frequency of torsional mode depending more strongly on $E$, since its second term in the bracket is not negligible, i.e. $4Et^2/(\sigma w^2)=3.58$ in our case. However, the torsional mode frequency of a corrugating string cannot be modelled properly with the above equation, thus we use FEM simulation instead to obtain a good fitting to the data. This alternative method allows us to measure $E$ of the material with high accuracy ($\Delta E<1$ GPa). Between the simulation results obtained with $E=186$ GPa and the data, the deviation $\Delta f_{tors}/f_{tors} = (f_{tors,data} - f_{tors, fit})/f_{tors,data}$ is less than 0.2\% for all torsional modes. All in all, the alternative procedure we demonstrate above allows us to determine the relevant mechanical properties of a-SiCR2 on fused silica.

\begin{figure*}[ht]
\centering
\includegraphics[width=1\textwidth]{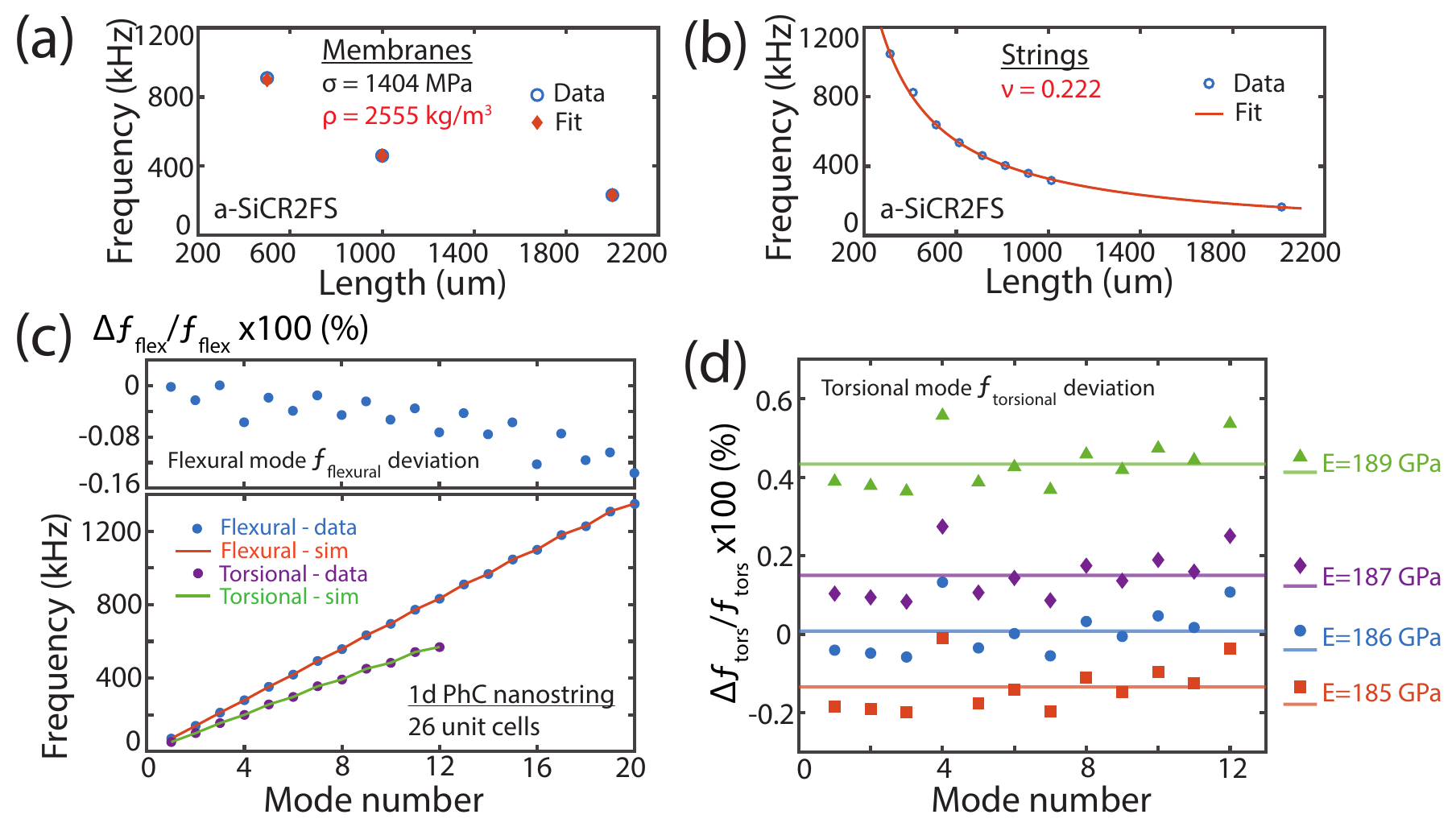}
\caption{Characterization of mechanical properties of a-SiCR2 thin film on fused silica substrate with alternative methods. (a) FEM simulation results (orange) to fit the measured data (blue circle) of the membranes with large undercut on the margins. (b) Measuring the Poisson ratio $\nu$ of the film by fitting (orange solid line) the data (blue circle) measured for strings. (c) The measured (blue/purple dots) and the FEM simulated (orange/green solid lines) resonant frequencies, of the flexural (up to 20th) and torsional modes (up to 12) of the 1d PnC nanostring with 26 unit cells are shown. Here the simulation results are obtained with the Young's modulus of 186 GPa. The flexural frequency deviation $\Delta f_{flex}/f_{flex}$ is shown in percentage on the top of the plot, and less than $0.16\%$ of the deviation is obtained. (d) Young's modulus $E$ determination using the torsional mode frequency of 1d PnC nanostrings. The scatter dots with different colors in the plot, represent the deviation between the measured data and the FEM outcome simulated with different $E$, whose values are shown on the right-side of the plot, i.e. $\Delta f_{tors}/f_{tors}$. The horizontal lines with various colors show the average values of the deviations of the scatter dots correspond to the same $E$. With this method, $E$ can be determined with high accuracy.}
\label{AMreply_SiCR2onFS}
\end{figure*}

\subsection*{A6. Plotting the frequency dependence on device lengths with different scales}

Different plots show different features in our case. Linear plots are clearer for the readers to comprehend the physical dimensions of the devices. Plots with 1/L and 1/L$^2$ can show the higher frequency data points clearer, but its clarity about the scale of the devices might be a bit lower for the readers, they are plotted in Figure \ref{AMreply_InverseScaleFit}; Alternatively log scale plot can show clearly the length dependence of different devices, and show the physical dimensions of the devices easily. We plot them in Figure \ref{AMreply_LogScaleFit}.\newline

\begin{figure*}[h]
\centering
\includegraphics[width=1\textwidth]{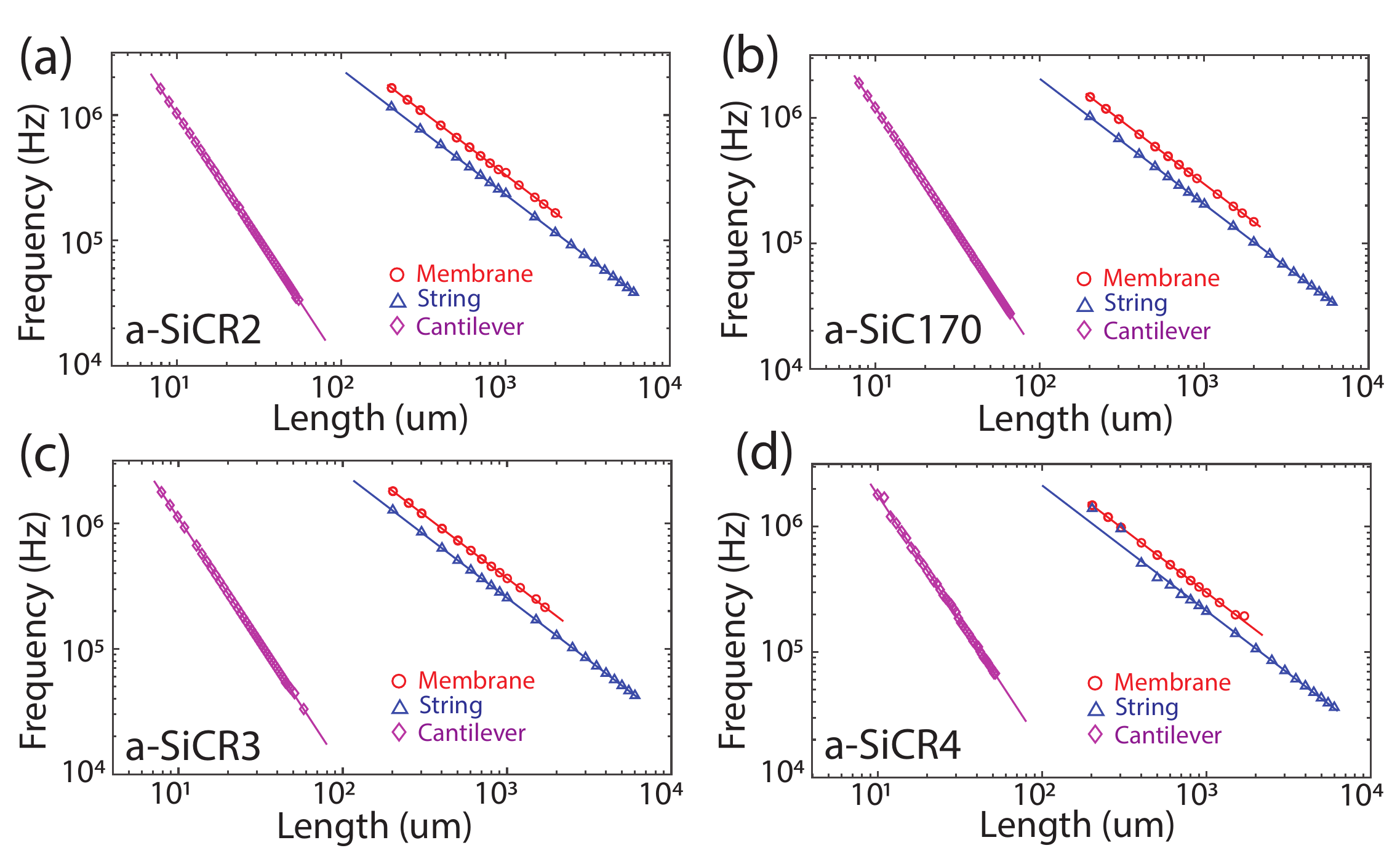}
\caption{Frequency as a function of length, plotted in log scale for (a) a-SiCR2 (b) a-SiC170. (c) a-SiCR3. (d) a-SiCR4 thin films. Frequency data for membranes (red circle), strings (blue upward triangular), and cantilevers (purple diamond) is shown, along with the fittings (red, blue, and purple solid lines, respectively).}
\label{AMreply_LogScaleFit}
\end{figure*}

\begin{figure*}
\centering
\includegraphics[width=1\textwidth]{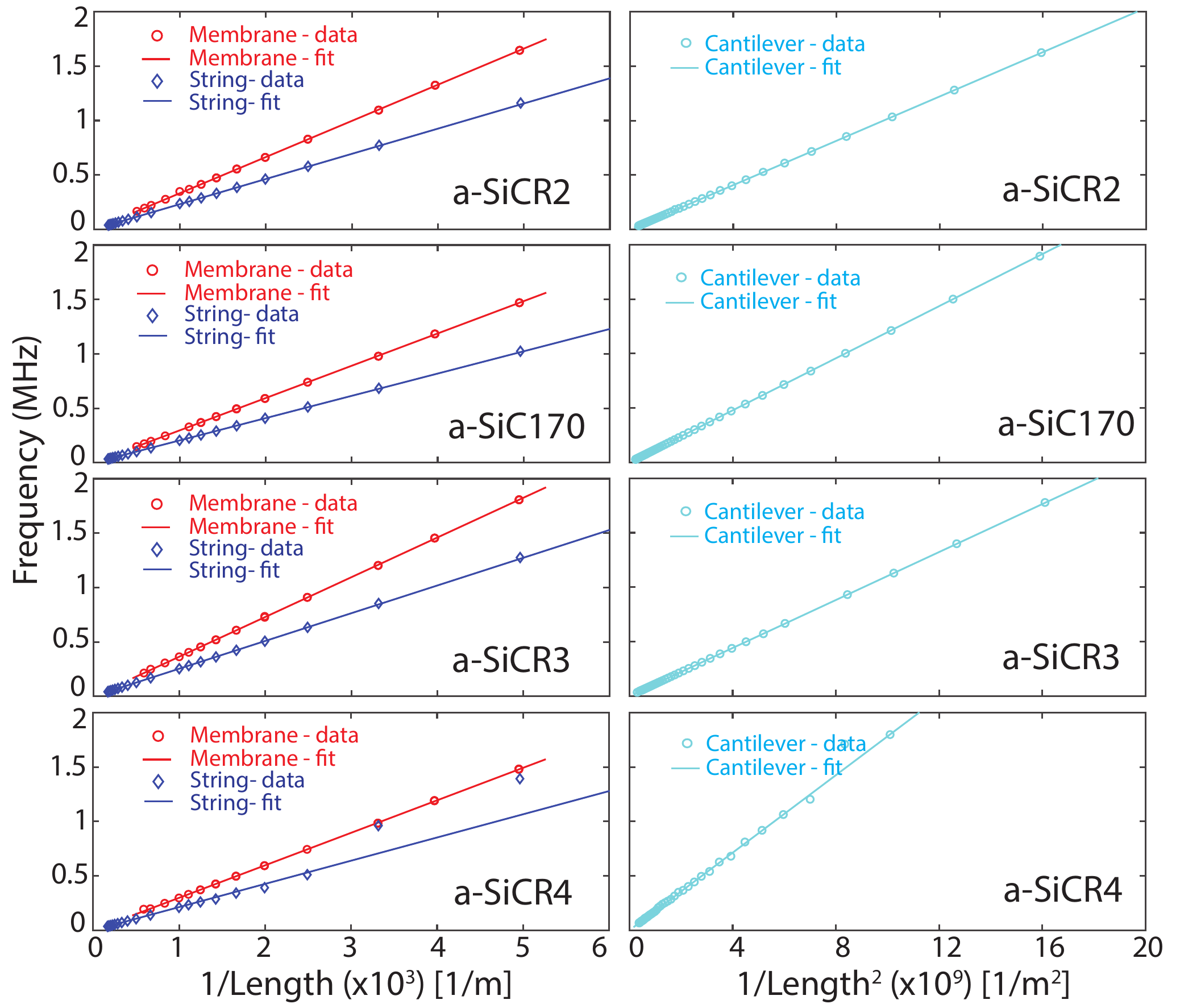}
\caption{Frequency as a function of 1/length for membranes and strings (plots on the left side),  and as a function of 1/length$^2$ for cantilevers (plots on the right side), are shown from top to bottom for a-SiCR2/170/R3/R4 thin films. Frequency data for membranes (red circle), strings (blue diamond), and cantilevers (cyan circle) is shown, along with the fittings with colors red, blue, and cyan solid lines, respectively.}
\label{AMreply_InverseScaleFit}
\end{figure*}

\newpage
$\ $
\newpage

\section*{Supporting Information \textbf{(B)}. Validation of the resonance method with COMSOL}

\begin{figure*}[h!]
\centering
\includegraphics[width=1\textwidth]{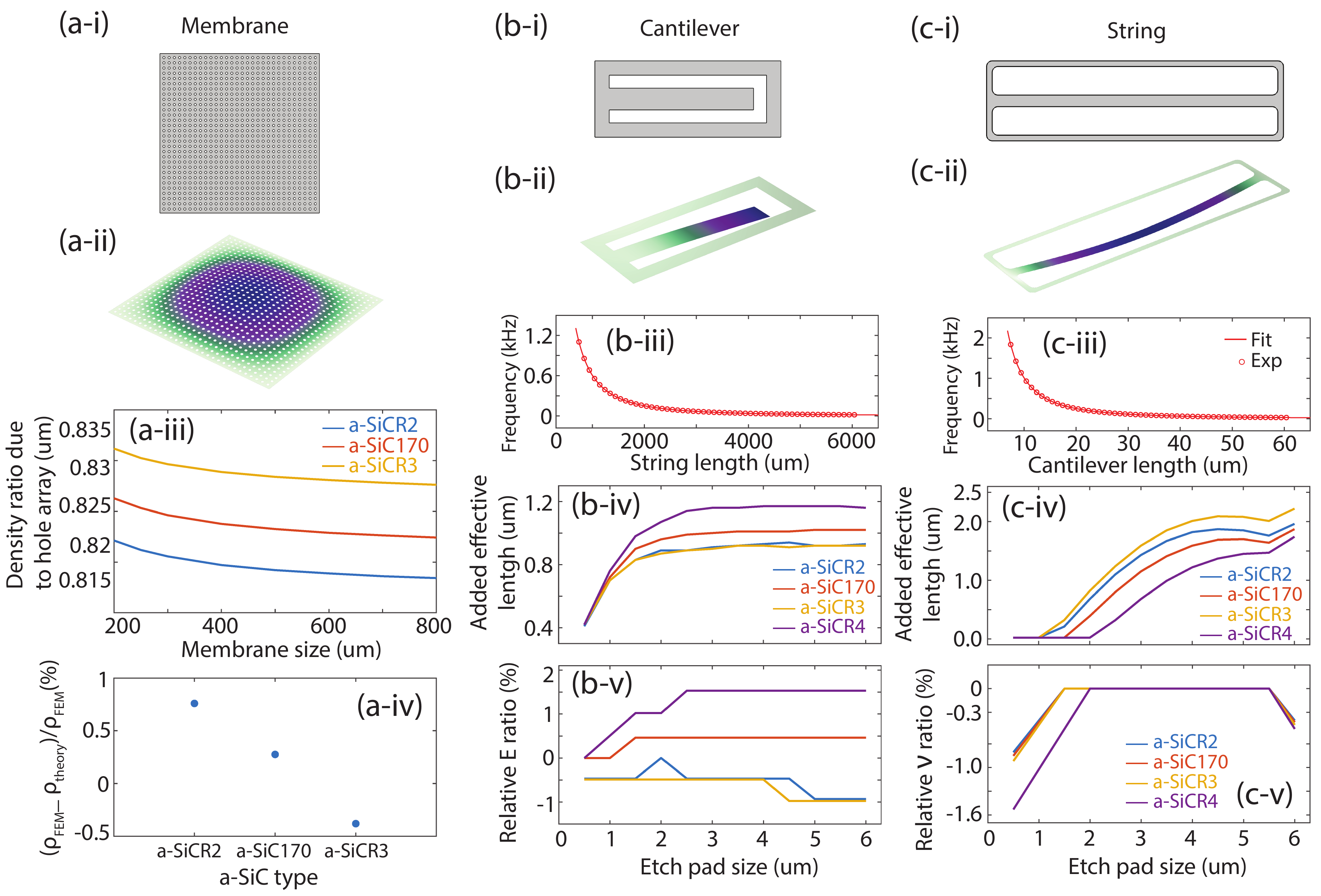}
\caption{Validation of using the resonant frequencies of membranes, strings and cantilevers to determine the density, Poisson ratio and Young's modulus of a thin film, respectively. The simulations are performed with finite element method (FEM) using COMSOL. In (a,b,c,-i and -ii), the geometrical shapes and mode shapes of their fundamental modes are illustrated. (a-iii) The influence on the effective density of the thin films due to the arrays of release holes, with hole radius 1.5 um and center separation 3.5 um. The sizes of the squared membranes are checked from 200 to 800 um, and the effective density values converge to 0.822 or close-by values in presence of the holes. (a-iv) For different a-SiC recipes, the density difference determined with FEM and the analytical formula is shown, which is less than 1\% by using the ratio value 0.822. (b-iii) and (c-iii): Using the fundamental mode analytical formula of the string and cantilever (solid line), to fit their FEM simulated results (red dot). (b-iv) and (c-iv): Studying the effect of the overhangs on the sides of the cantilevers and strings generated during the undercut process, by fitting the FEM results given various undercut sizes, with the analytical formulas with effective lengths ($L_{eff}$) of the geometries. The comparisons show that both for cantilevers and strings, the influence on the added effective lengths will saturate to some values smaller than the real overhang sizes, ensuring the stability of the method. (b-v) and (c-v): Comparing the Young's modulus $E$ and Poisson ratio $\nu$ obtained with both the FEM simulations and analytical formulas, as a function of the undercut pad sizes. The ratios of error between the two methods are less than 1.5\% for determining $E$ and less than 1.6\% for determining $\nu$, validating the accuracy on identifying both mechanical properties with this method based on measuring the resonant frequencies of cantilevers and strings.}
\label{MechProValid}
\end{figure*}

\newpage

\section*{Supporting Information \textbf{(C)}: Theory of dissipation dilution, and dilution factors of 1d PnC nanostings}

In this work, the intrinsic quality factor $Q_{int}$ is determined by measuring the mechanical quality factors $Q_D$ of PnC nanostrings \cite{Fedorov2019}

\begin{equation}
    Q_D = D \cdot Q_{int},
\end{equation}

where the dilution factors $D$ are calculated numerically, it depends on the various mechanical properties of the materials, as well as the geometry of the resonator. For a string-like resonator with thickness $t$ and length $L$, the dilution factor has the factor
\begin{equation}
    D_n = \frac{1}{2\lambda + \pi^2 n^2 \lambda^2},
\end{equation}

where $n$ is the mode number of the resonator and $\lambda$ is defined as
\begin{equation}
    \lambda = \frac{h}{L}\sqrt{\frac{E}{12\sigma}}.
\end{equation}

In order to further investigate its applicability to fabricate high-Q nanomechanical resonators, we need to identify the intrinsic quality factor $Q_0$ of a-SiC thin films, which is most accurately by experimentally measuring Q factor of geometrically strain-engineered resonators whose external loss mechanisms are eliminated and dissipation dilution factor $D_Q$ is well defined, leading to an expected intrinsic Q factor $Q_0=Q/D_Q$. To perform such experiments, we fabricate a series of uniformly corrugated high-aspect-ratio phononic crystal (PnC) nanostrings of length 4 mm, whose unit-cell lengths $L_{uc}$ together with defect lengths $L_{def}$ in the middle are varied (Figure 4(a)), leading to PnC nanostrings of unit-cell numbers from 20 to 44. The widths of the wide and narrow parts of the nanostrings are 3 um and 1 um respectively. With higher unit-cell number or shorter defect length, the PnC nanostring has a defect mode located in a phononic bandgap at higher frequency, the example of PnC nanostring with 20 unit cells is shown in Figure 4(b). The vibration amplitudes of the nanostrings as a function of frequency are acquired with a custom balanced homodyne detection interferometer under a vacuum level of $4\times 10^{-9}$ mbar, with them the engineered phononic bandgaps of the PnC nanostrings are identify and the defect modes inside are confirmed. With the ringdown method, the defect mode Q factors of the PnC nanostrings are measured, see Figure 4(d). Using finite element method (FEM) simulation, the dilution factor $D_Q$ of each PnC nanostring geometry can be numerically calculated, together with the Q factors of the corresponding nanostring measured experimentally, the intrinsic Q factor $Q_0$ of a-SiC thin films are determined, as shown in Figure 4(e-f). We employ PnC nanostrings for intrinsic Q factor identification instead of other geometries such as membranes \cite{Villanueva2014} or normal strings \cite{Manjeshwar2022} as shown for other works, since their FEM simulated $D_Q$ are much less dependent on the meshing at the clamping edges, as well as the measured Q factors of the localized defect mode do not rely on how the resonators link to the substrate. The intrinsic loss $Q_0$ of a-SiC films can be attributed to the volume loss $Q_{vol}$ and the surface loss $Q_{surf}$, i.e. $Q_0 = (1/Q_{vol} + 1/Q_{surf}\cdot t)^{-1}$, for our thin film resonators the low surface-to-volume ratio allows us to set $Q_{vol}$ to be the same as the the one of LPCVD a-SiN, i.e. 28000 (see \textbf{Figure \ref{SfigExpQ}(f)} for more detailed), and $Q_{surf}$ is proportional to the thickness $t$ of the corresponding film, which we compare with the one of LPCVD a-SiN for clarification, i.e. $Q_{surf}^{SiC} = x\cdot Q_{surf}^{SiN}$, where $x$ is the ratio between the two surface loss and $Q_{surf}^{SiN}=6900\cdot t/100[nm]$ is the surface loss of a-SiN. 

We can analytically calculate the Q factor of a phononic crystal (PnC) nanostring because its losses are dominated by bending-losses from the material and does not include clamping losses to the substrate. However, simple double-clamped strings have modes which couple directly to the substrate, and thus the mechanical loss is now dependent on both the nanostring material and the substrate, which make difficult to use quality factor of simple nanostrings to intrinsic Q of the material ($Q_0)$. As is shown in \cite{Manjeshwar2022}, for simple doubly-clamped string, other external loss mechanism will influence the accuracy on measuring $Q_0$. By using the 1d PnC nanostring instead, our mode of interest has no coupling with the substrate. Moreover, using phononic crystal nanostrings we can greatly enhance the dilution factor $D_n$, and thus a small difference in $Q_0$ will be amplified into a noticeable difference in the quality factor $Q=D_n \cdot Q_0$.The Q factor dependence on frequency/length, for a specific material with a constant $Q_0$, is equivalent to checking the dilution factor $D_n$ dependence on frequency/length.

\begin{figure*}
\centering
\includegraphics[width=1\textwidth]{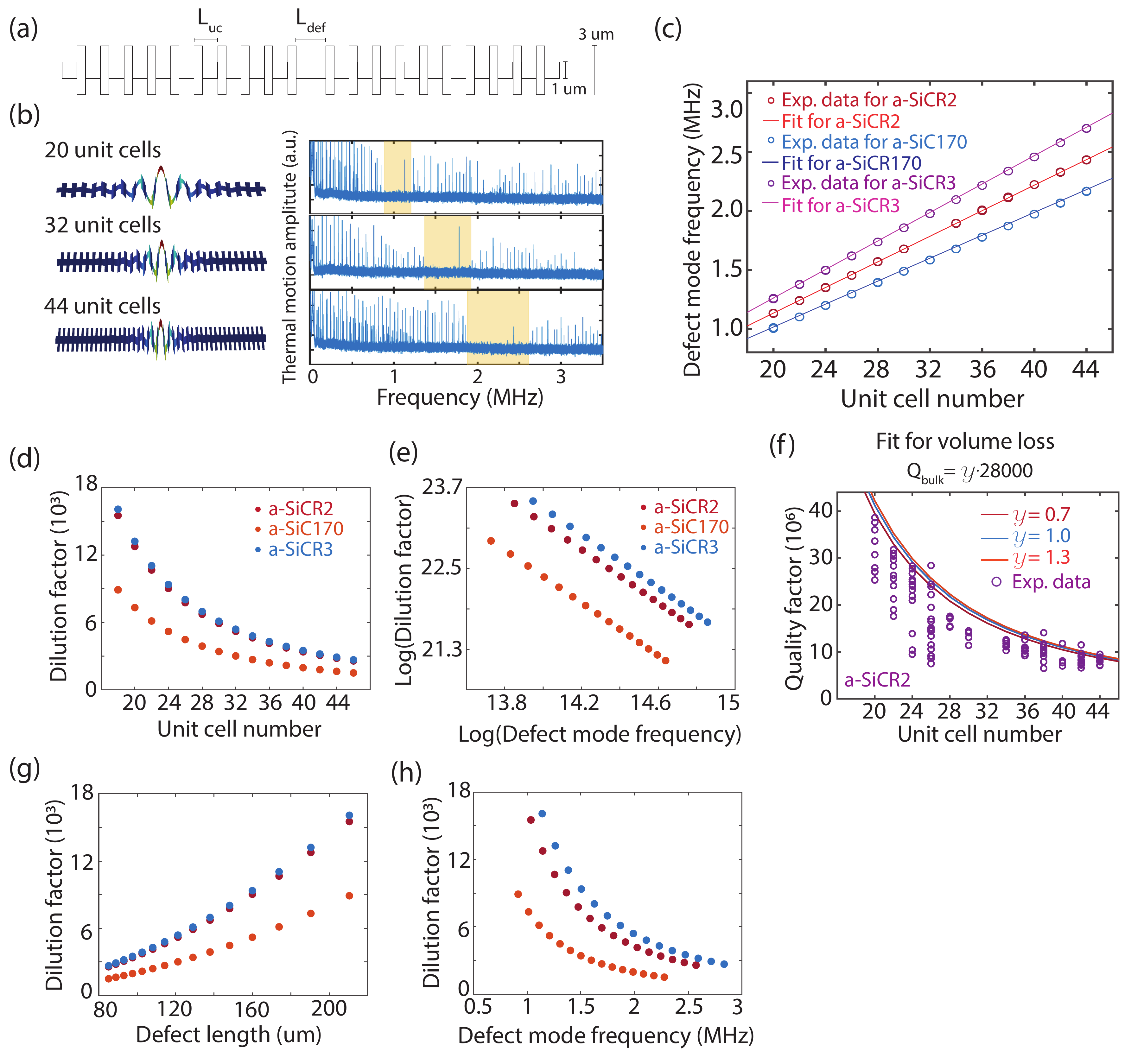}
\caption{1d phononic crystal (PnC) nanostrings for checking the intrinsic quality factors (Q factor) of a-SiC films. (a) geometry of a 20 unit-cell (uc) PnC nanostring. (b) Mode shapes and frequency spectrums of 20/32/44 uc PnC nanostrings. (c) Resonant frequencies of the defect/localized modes of PnC nanostrings with 20-44 uc nanostrings, the measured data (hollow dot) are fitted linearly (solid line). (d) and (e): Dilution factor $D_n$ as a function of PnC nanostrings with different numbers of unit cells, in linear and log scales respectively. (f) Q factors fitted with different bulk intrinsic quality factor, the small deviation between the adjacent lines indicates thicker thin film resonators and more accurate measurement data are required to get proper values for this value.}
\label{SfigExpQ}
\end{figure*}

\newpage
$\ $
\newpage
\section*{Supporting Information \textbf{(D)}: More data on ringdown measurement of 1d PnC nanostrings, and intrinsic Q factor characterization of a-SiC thin films}

\begin{figure*}[h]
\centering
\includegraphics[width=0.9\textwidth]{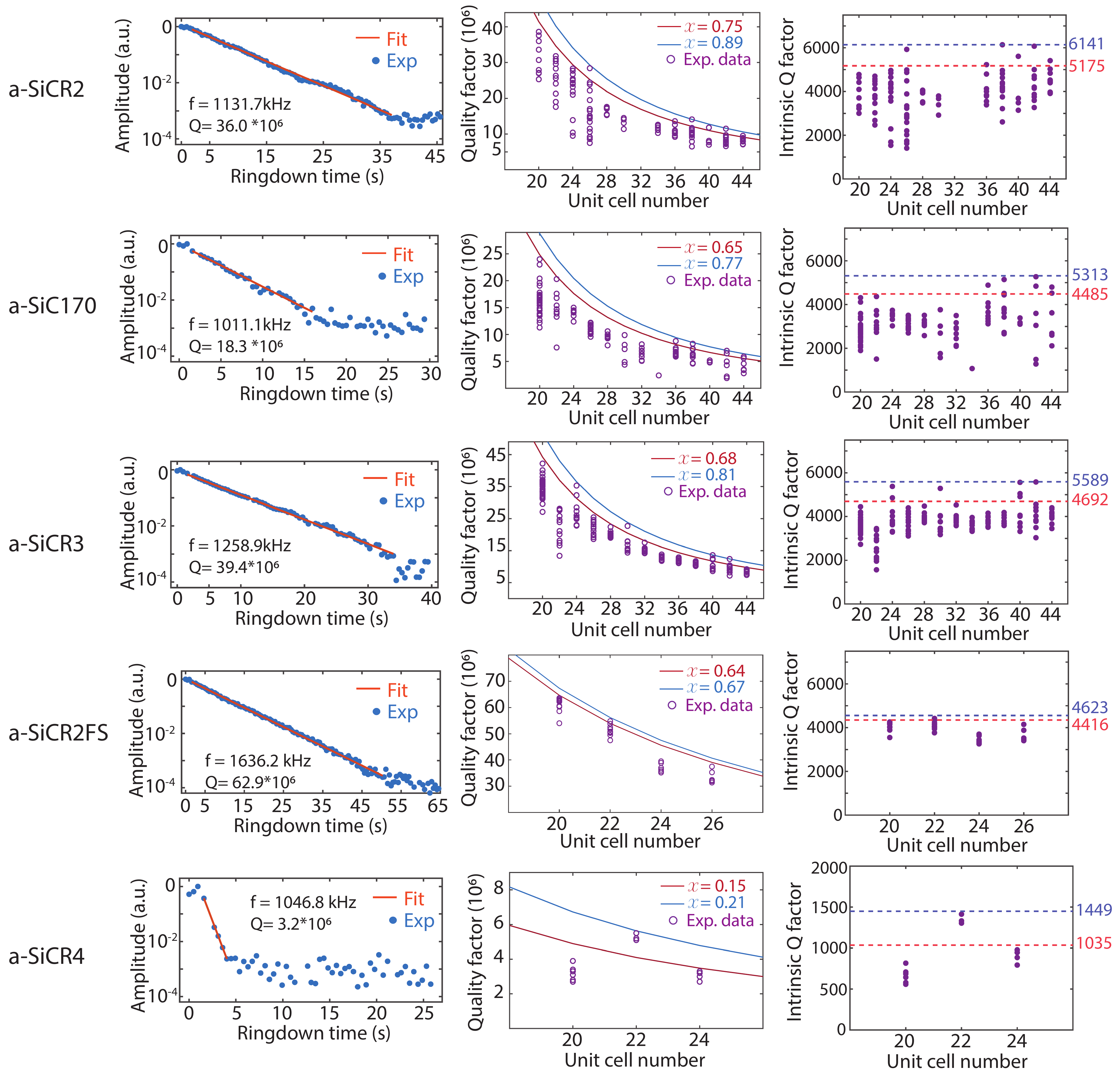}
\caption{More data on intrinsic quality factors of different LPCVD a-SiC thin films, from top to bottom rows: a-SiCR2, a-SiC170, a-SiCR3, a-SiCR2onFS, a-SiCR4. Plots on left, middle and right columns, show the ringdown data (left), fitting of intrinsic Q factors with measured Q factor before (middle) and after (right) divided by the dilution factors. The measured intrinsic Q factors are shown in \textbf{Table 1}.}
\end{figure*}

\newpage
\section*{Supporting Information \textbf{(E)}: Machine learning technique for designing high-Q resonator}

\begin{figure*}[h]
\centering
\includegraphics[width=0.6\textwidth]{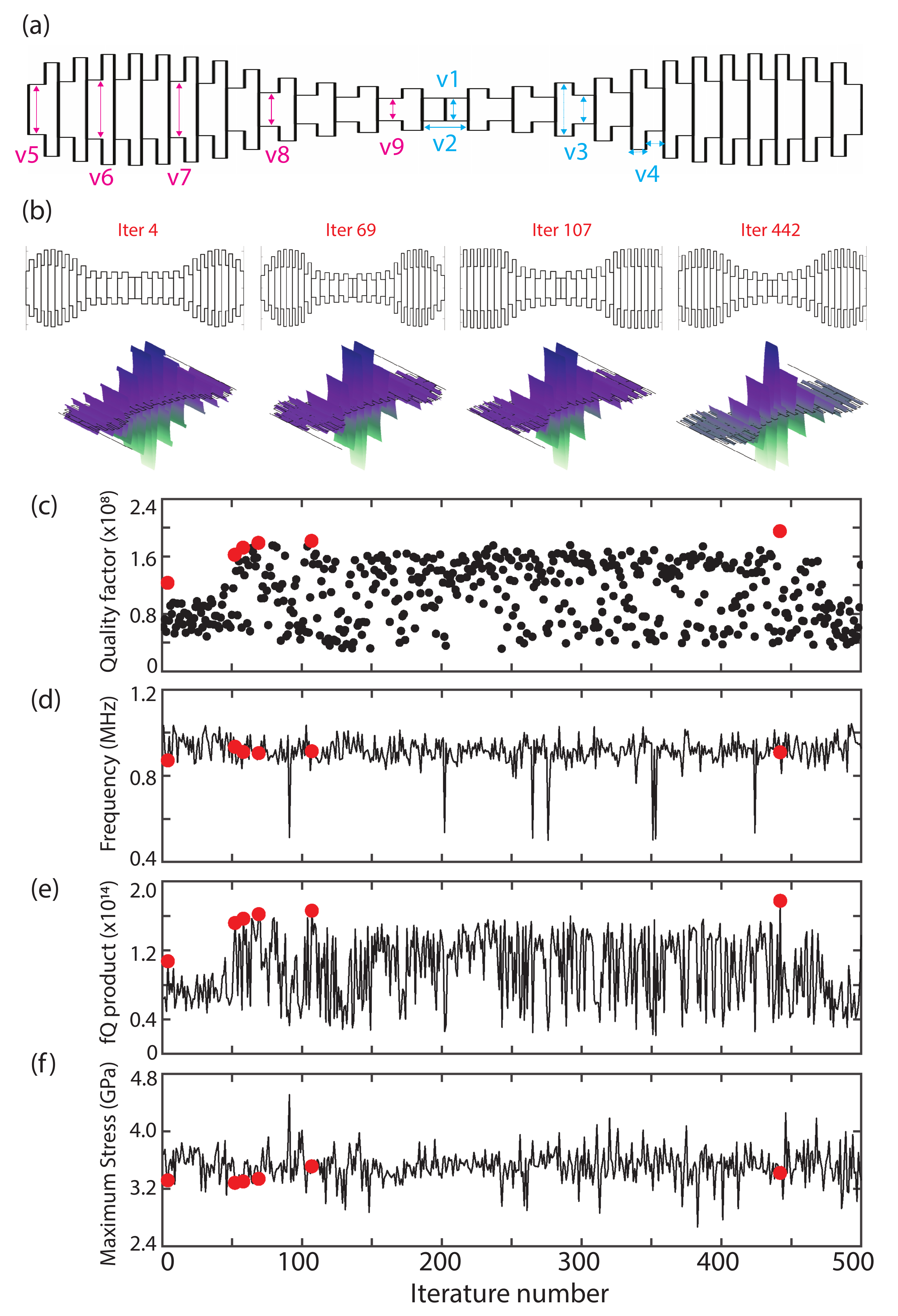}
\caption{(a) Parameters for optimization. (b) The optimized designs up to a certain optimization cycle are marked with red dots, these 4 designs' geometries and localized mode shapes. (c) Optimized Q factors results from intermediate steps, here we assume the intrinsic Q factor of a-SiCR2 to be 5175/100nm. Together with its Q factor the other properties of the designed resonator can be obtained, here its resonant frequency, frequency (f) times quality factor (Q) product, and the maximum tensile stress on the nanostring are shown in (d), (e) and (f), respectively. }
\end{figure*}

Bayesian optimization for ultra-high Q a-SiC nanostring made with a-SiCR2. The nanostring has a total length of 6 mm and the number of unit cells is fixed to 24. In the Bayesian optimization algorithm, the Q factor of the nanomechanical resonators is being optimized. 9 design parameters are setup for the optimization. The iteration steps are in total 500. Among the parameters, v1 is the defect width, constrained between 1 to 3 um; v2 is the defect length, constrained between 50 to 500 um; v3 is unitcell width ratio, constrained between 1:1.5 to 1:3; v4 is unitcell length ratio, constrained between 1:3 to 3:1; v5 to v9 are the widths of the unitcells' thin parts, the width parameters defining the tapering shape, constrained between 1 to 3 um. Lengths of the unitcells was determined considering bandgap frequency matching condition, once the set of each unitcell's width is defined \cite{Ghadimi2018}. One can find that the mode shapes are getting more and more confined from the clamping points to minimize the clamping loss. From the fifth optimized result (Iter 107) to the final optimized result (Iter 442), the geometry is changing from a uniformly corrugated design on the edge to a non-uniformly corrugated one, this interesting found might lead to interesting perspectives in designing the 1d PnC nanostring in the future. Interesting to note that the highest Q factor design doesn't coincide with the design with the highest tensile stress.

\section*{Supporting Information \textbf{(F)}: Thin a-SiC resonator}

\subsection*{F1. SEM pictures of nano-thick a-SiC films}
\begin{figure*}[h!]
\centering
\includegraphics[width=0.9\textwidth]{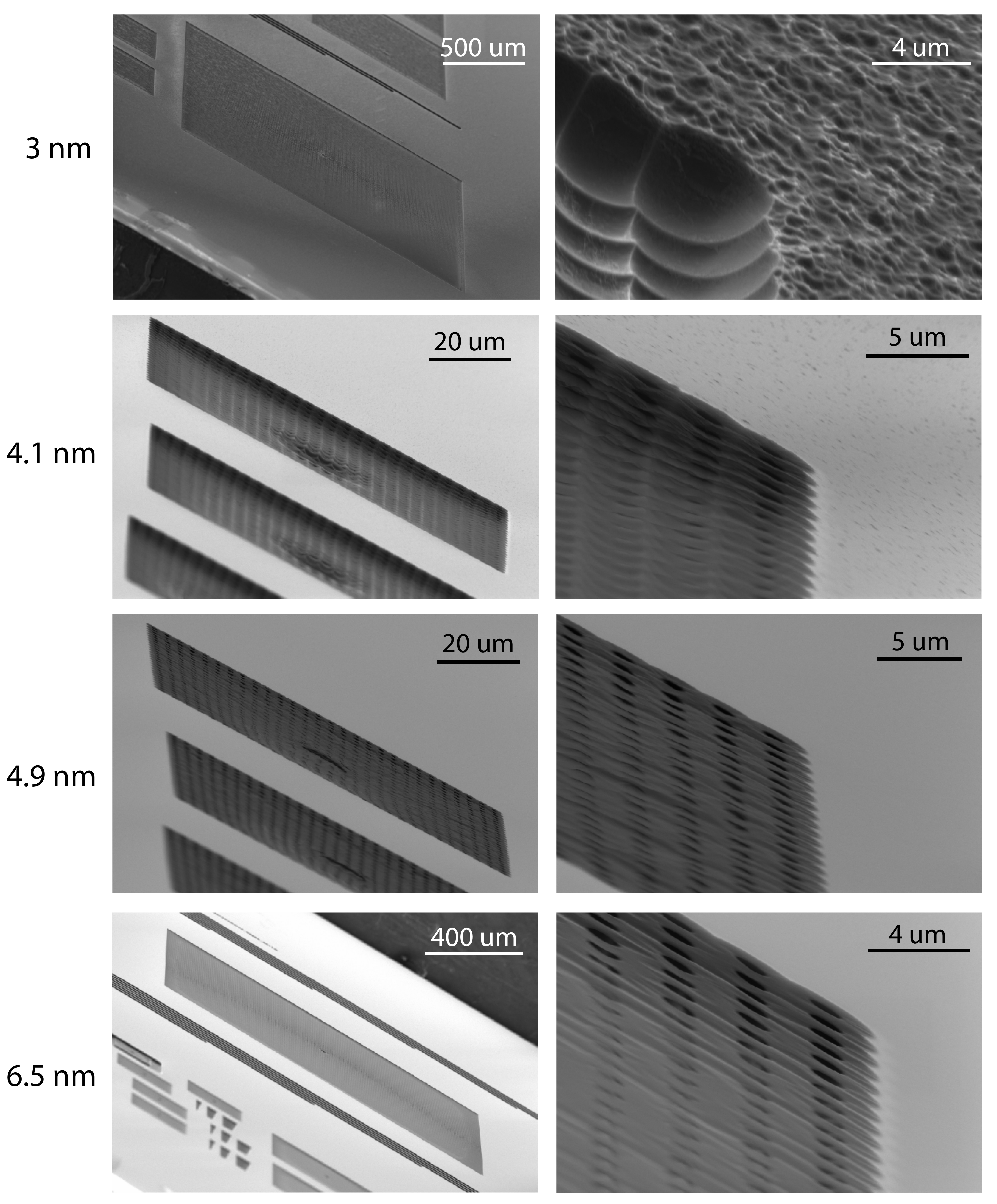}
\caption{LPCVD a-SiC squared membranes fabricated with a-SiCR2 thin films polished with the ion beam etcher (SCIA Ion Mill 150). From top to bottom, the a-SiCR2 thin film is thinned down from 79 nm to 3/4.1/4.9/6.5 nm, respectively. The SEM images on the left are the zoom out image of the resonator, while the images on the right are the zoom in images of the top-right corner of the images on the left-hand side. One can observe that the down to 4.9 nm, the LPCVD a-SiC films are continuous and suspended membranes can be fabricated, the pinholes start to appear when the film thickness is polished down to 4.1 nm, and a rough surface is presented at the film thickness of 3 nm. The pinholes appearance on the first few nanometers put a limitation on the ultimate a-SiC resonator thickness one can work with.}
\end{figure*}

\begin{figure*}[h!]
\centering
\includegraphics[width=1\textwidth]{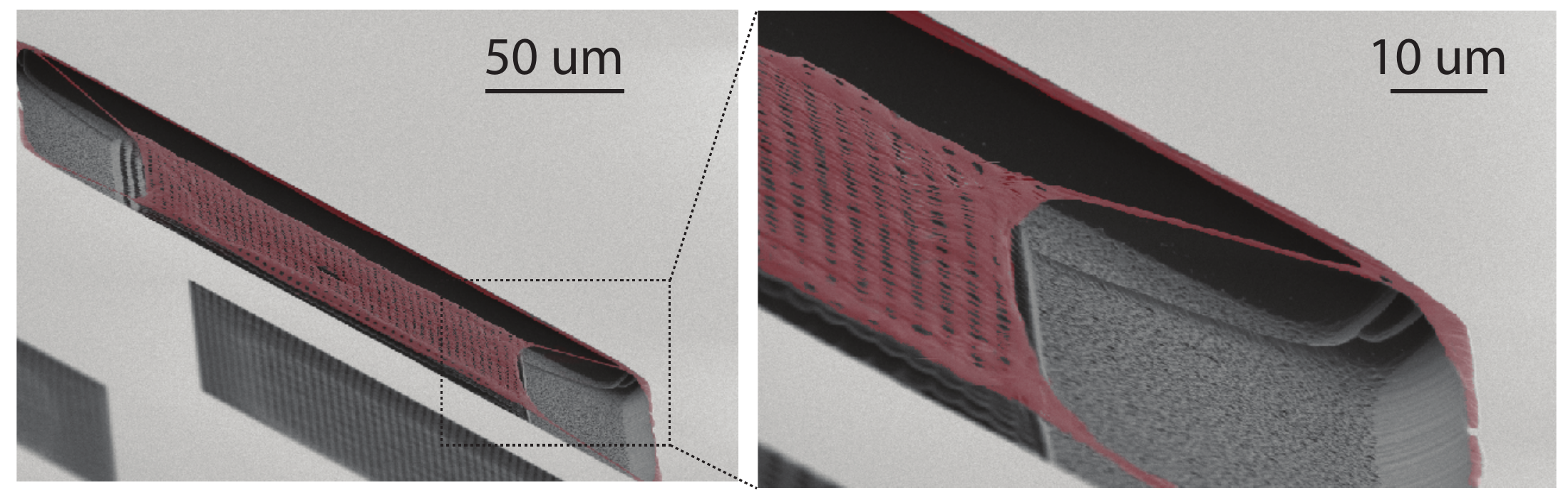}
\caption{LPCVD a-SiC trampoline resonator fabricated with a-SiCR2 thin films of length 250 um and thickness 4.9 nm, polished with the ion beam etcher (SCIA Ion Mill 150). The trampoline resonator remains suspended after undercut. No pinholes are observed on the thin film. The aspect ratio of this trampoline resonator is comparable to the best ones a-Si$_3$N$_4$ resonators have achieved.}
\end{figure*}

Both the force sensitivity and force responsivity of nanomechanical sensors are limited by the minimum thickness achievable for the continuous films. For thin film sensors with high intrinsic tensile stress , higher aspect-ratio (length/thickness) can increase the feasibility of higher Q factor, thus higher force sensitivity due to lower noise level; While for the ones with low film stress, thinner films lead to lower stiffness, therefore larger deflection under given forces, or higher force responsivity. Unlike 2D materials such as graphene, which use the bottom-up approach to construct ultra-thin suspended resonators (e.g. growing the thin film layer by layer atomically), engineerable thin films such as a-SiN and s-Si usually use the top-down approach to get ultra-thin resonators. As respectively shown in \cite{Hoj2021} and \cite{Beccari2022}, in order to fabricate high-aspect-ratio (HAR) resonators with thicknesses down to 12nm SiN and 15nm (sSi) thin films, encapsulating layers are required to hold up as well as to protect the thin SiN and sSi structures, which complicate the fabrication processes and may even introduce contamination. Compared to the above materials commonly used for nanomechanics, a-SiC thin films can be fabricated into thinner resonators thanks to its superior chemical inertness. As shown previously, even with LPCVD, which is considered to deposit continuous and conformal films, cannot avoid pinholes on the films while depositing the first nanometers. In order to find out the minimum thickness achievale for a-SiC thin film, the ion beam etcher (SCIA Ion Mill 150) is used to thin down the thicker and more uniform GFR2 films down to the desire thickness. The ion beam etcher, compared to other techniques, such as reactive plasma etching, is preferred to get ultra-thin films down to sub-nanometer accuracy, with a physical sputtering process applicable to a wide range of materials called ion beam milling, which avoid surface contamination and local charge accumulation that might hinder to obtain the desired conformal thin film accurately. The GFE2 films with initial thickness 71nm are thinned down with a beam voltage of 120V (kinetic energy 120eV per ion) and an incident angle of 4 degrees sheering, resulting in an etch rate of 7 nm/h. By suspending resonators with polished film thicknesses from 3 to 6 nm on silicon substrate, we found that minimum continuous smooth LPCVD a-SiC GFR2 film achievable is between 4.1 and 4.9 nm. Worth to note that buckling patterns are observed on the trampoline resonator, due to the facts that: 1. the thinner and larger the resonator is, the smaller stress gradients in both in-plane and out-of-plane directions are required in order to keep the resonator flat \cite{Beccari2022,Bereyhi2019}; 2. the first several nanometers of the a-SiC film is with less tensile or even compressive stress due to defects at the material interface at the start of the deposition, which are also shown in \cite{Hoj2021,Romero2020}.

The surface topography of the a-SiCR2 thin film polished down to 3nm, is measured with AFM shown in \textbf{SI. G(d)}, which is only half the RMS compared to the one prior to being polished.

\newpage
\ 
\newpage
\ 
\newpage

\subsection*{F2. XPS measurement on the nano-thick a-SiC films}
Thank you for bringing this up. Indeed about a nanometer of the film will be native oxide, thus for the nano- a-SiC films (4.9 to 6 nm), it is necessary to show it still consist of silicon and carbon chemical components and Si-C bonds. We performed XPS measurement on three different spots: (1) 6-nm-thick a-SiCR2 film without experiencing SF$_6$ isotropic etching, (2) 4.9-nm-thick a-SiCR2 film after SF$_6$ etching, (3) and 4.9-nm-thick a-SiCR2 membrane (1.5x1.5 mm$^2$) after SF$_6$ etching. The results of the XPS measurement is shown in Figure \ref{AMreply_XPS}. From the spectrum shown in Figure \ref{AMreply_XPS}(a), peaks correspond to silicon, carbon and oxygen components are measured for all three spot. Fluorine peak is found for films experienced the SF$_6$ cryogenic etching, this was also observed in \cite{pereira2009situ}.

In Figure \ref{AMreply_XPS}(b) and (c), detailed XPS spectra around Si(2p) and C(1s) are shown, respectively. In Figure \ref{AMreply_XPS}(b), the XPS spectrum of 6-nm-thick film without SF$_6$ cryo etching is fitted (dash purple line), with Si-C bond (101.2eV, purple solid line), Si-O-C/Si-O bond (102.7eV, cyan solid line) and SiO$_2$ bond (103.7eV, yellow solid line) \cite{kaur2015selective}. The slight difference between the spectrum of the films before and after etching, may come from Si-F and/or C-F bonds. In Figure \ref{AMreply_XPS}(c), the XPS spectrum of 6-nm-thick film without SF$_6$ cryo etching is fitted (dash purple line), with Si-C bond (285.3eV, purple solid line), C-O bond (287.6, cyan solid line) and C-C bond (284.3eV, yellow solid line) \cite{mckenna2011synthesis}. From the XPS spectrum, the Si-C bond is identified in all three measurement spots, indicating that nano-thick SiC films still consist if a-SiC. This plot is also added to the SI for a better insight of the nano-thick a-SiC thin film.\newline

\begin{figure*}[ht]
\centering
\includegraphics[width=1\textwidth]{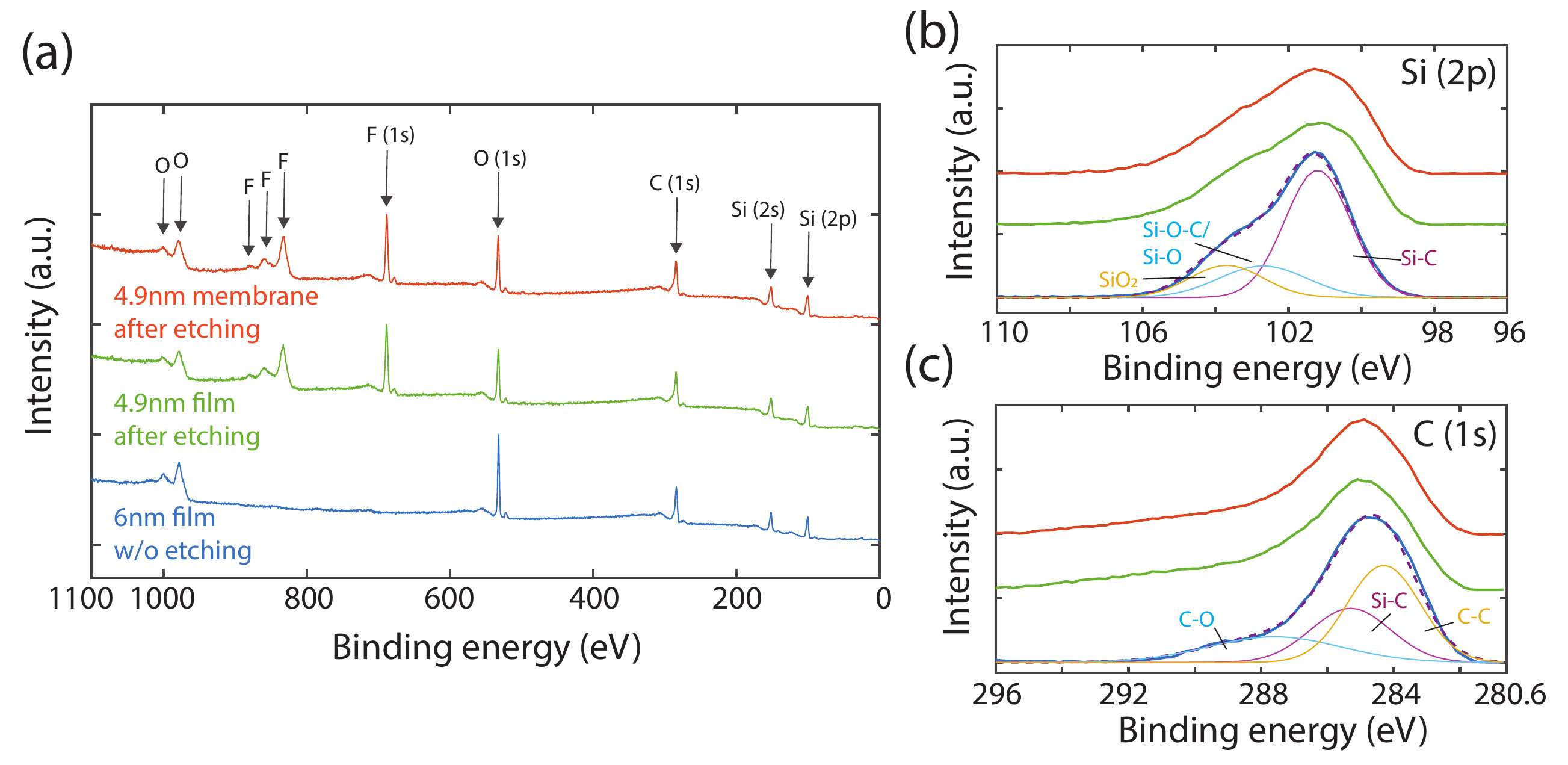}
\caption{Chemical composition characterization of the nano-thick a-SiCR2 films with X-ray photoelectron spectroscopy (XPS). (a) XPS spectrum of 6-nm-thick a-SiCR2 film without SF$_6$ cryo etching (blue), 4.9-nm-thick a-SiCR2 film after SF$_6$ cryo etching (green), and 4.9-nm-thick a-SiCR2 membrane (1.5x1.5 mm$^2$) after SF$_6$ cryo etching (red). This color code is used also for (b-c). Compositions such as silicon, carbon, and oxygen are measured for all three thin films. For films experienced SF$_6$ cryo etching, fluorine residual is found. Detailed XPS spectra around (b) the Si(2p) peak and (c) C(1s) peak. The XPS spectrum of 6-nm-thick film without SF$_6$ cryo etching (blue solid line) is fitted (dash purple line).}
\label{AMreply_XPS}
\end{figure*}

\newpage

\section*{Supporting Information \textbf{(G)}: Surface topography of a-SiC thin films with Atomic force microscopy}

Among all recipes, a-SiCR2 and a-SiCR3 have the lowest roughness. Yet one can still find large bright spots on a-SiCR3 films with the dark field image from the optical microscope, indicating that it is not as flat in a larger landscape. After polishing with the ion beam miller, a-SiCR2 with 3 nm is scanned, with only half the roughness compared to the one prior to polishing. Interesting to note that, with lower deposition pressure, a-SiC170 is rougher than a-SiCR2, which matches well with its lower intrinsic Q factor and lower fracture strength. The topography of a-SiCR4 is very rough, can be due to the re-crystallization of Si components in the thin film.

\begin{figure*}[h!]
\centering
\includegraphics[width=0.9\textwidth]{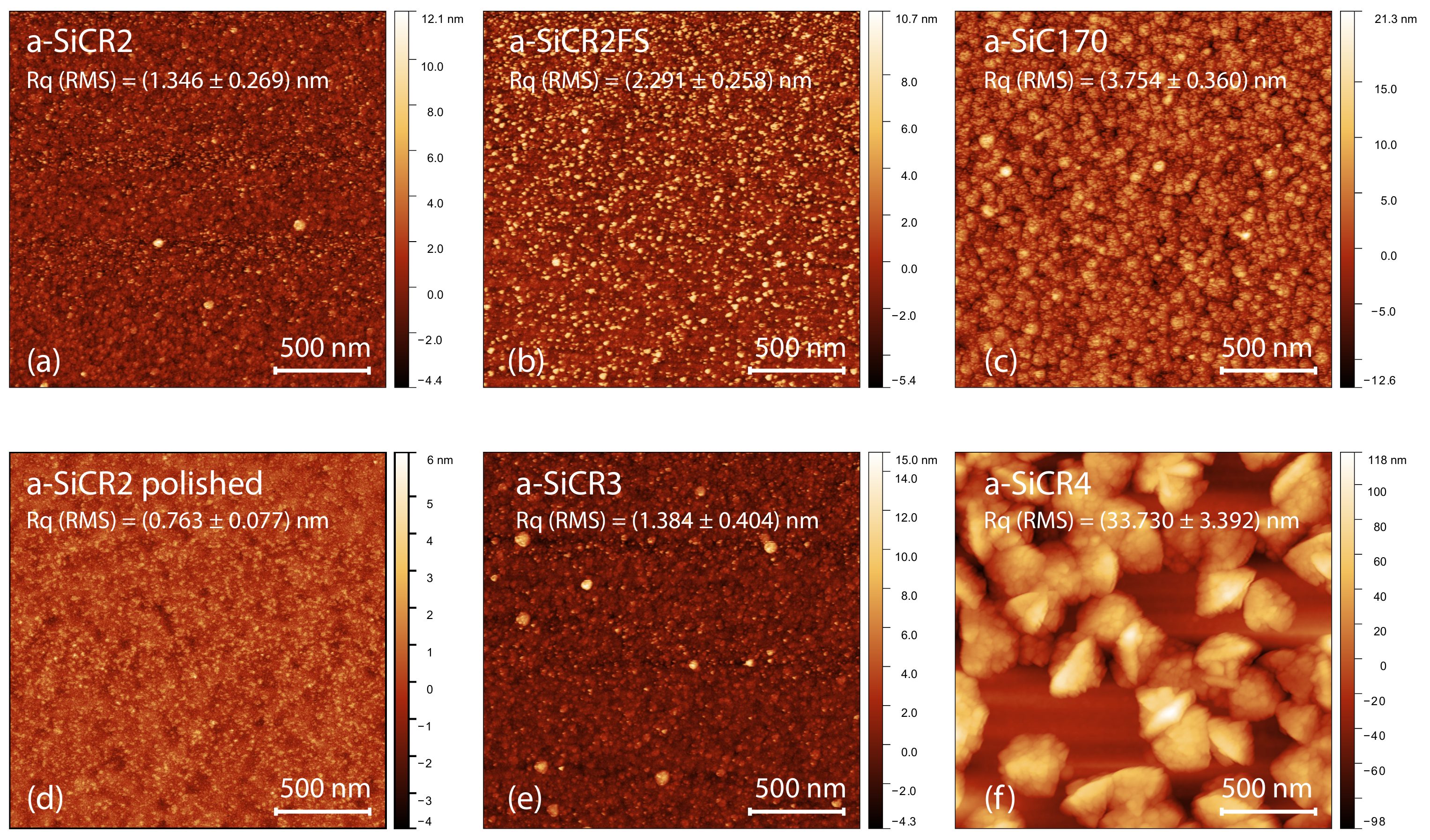}
\caption{Surface topography of LPCVD a-SiC thin films scanned with atomic force microscope (AFM). From left to right, the first row shows topography of (a) a-SiCR2, (b) a-SiCR2FS, (c) a-SiC170; the second row shows the topography of (d) a-SiCR2 being polished down to a thickness of 3 nm using the ion beam miller, (e) a-SiCR3, and (f) a-SiCR4.}
\end{figure*}

\begin{figure*}[h!]
\centering
\includegraphics[width=0.6\textwidth]{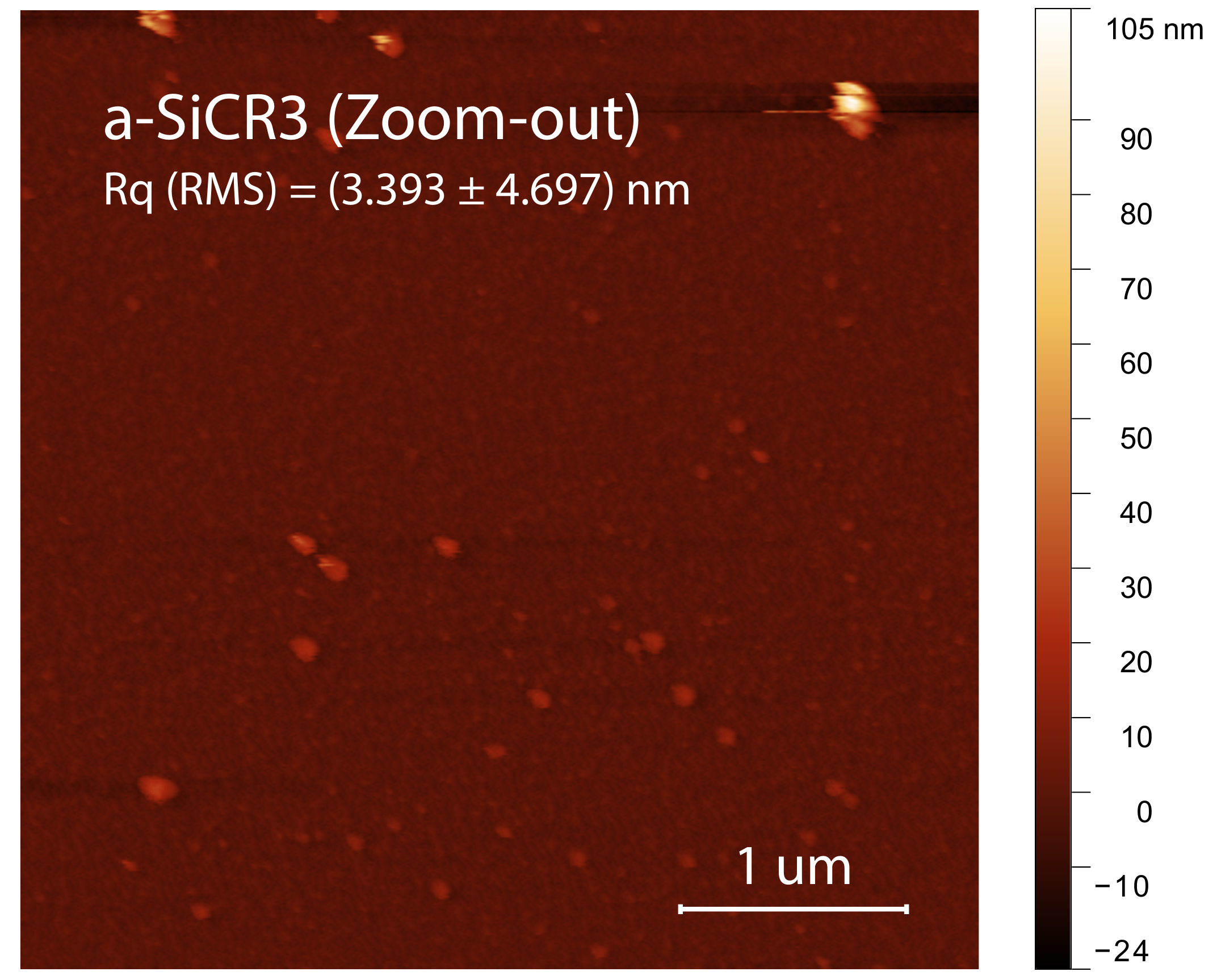}
\caption{Zoom-out Surface topography of a-a-SiCR3 thin film scanned with atomic force microscope (AFM).}
\end{figure*}

\newpage
$\ $
\newpage

\section*{Supporting Information \textbf{(H)}: Chemical composition characterization of the a-SiC films}

There is a necessity to perform a detailed examination and discussion on the film composition and chemical bond structure in relation to their mechanical properties. Here we have undertaken a series of meticulous analyses, including XRD, EDX, and Raman spectroscopy, to gain deeper insights into these aspects.

In order to make the characterization results more comparable, the a-SiCR2/170/R3/R4 thin films used for EDX and Raman spectroscopy, are of thicknesses 78/86/81/137 nm, respectively.

The XRD analyses (Fig \ref{AMreply2}a) revealed no discernible peaks related to crystal or poly-crystalline structures, underscoring the amorphous nature of the films \cite{yu2019third}. A noteworthy observation from our Raman spectroscopy (Fig \ref{AMreply2}c) is the correlation between higher strength and a lower ratio of Si-Si/C-C bonds.  This finding aligns with established knowledge \cite{MoranaPhD2015}; generally, C-C single bonds, with a bond dissociation energy of approximately 348 kJ/mol (3.61~eV), are stronger than Si-Si single bonds, which possess a bond dissociation energy of around 226 kJ/mol (2.34~eV) \cite{Cui2019}.


We have integrated these additional insights and discussions into the concluding sections of our manuscript, and believe this enriches the paper by not only providing readers with a nuanced understanding of the properties of these films but also offering intuitive guidance on reproducing and optimizing these films further. These additions to the conclusion aim to clarify the compositional attributes that confer the mechanical properties observed in our a-SiC films.

\begin{figure*}[ht]
\centering
\includegraphics[width=0.8\textwidth]{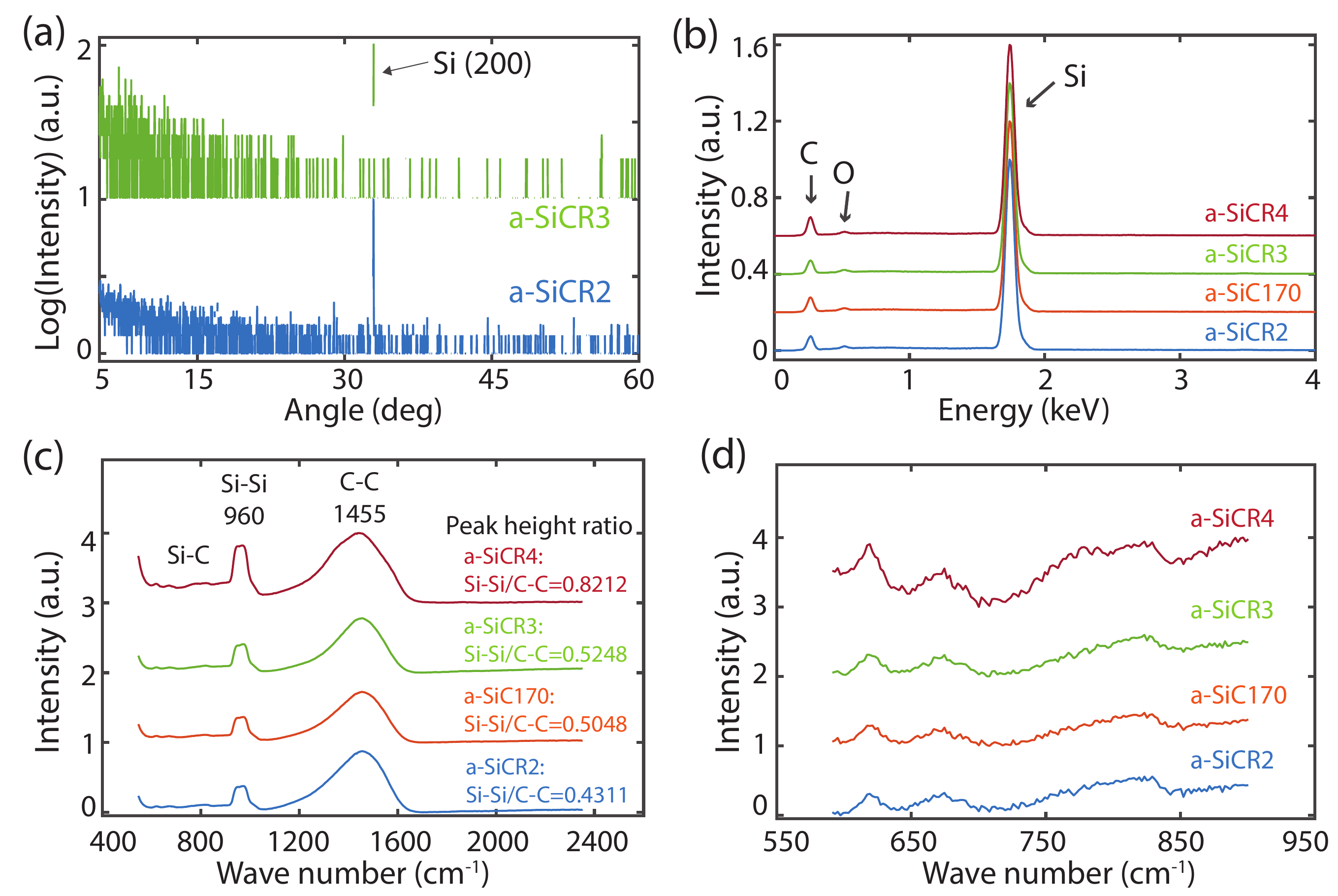}
\caption{Characterization of the chemical composition of a-SiC films. (a) X-Ray diffraction (XRD) spectrum to check the crystal structure of a-SiCR2 (blue) and a-SiCR3 (green) thin films studied in this work. The peaks at 32.95$^{\circ}$ can be assigned to Si(200). No additional signal peak can be identified on the rest of the spectrum. (b) Energy-dispersive X-ray (EDX) spectroscopy to measure the chemical composition of the a-SiC thin films, all of the four spectrum indicate the films are composed of carbon, silicon, and oxygen. (c) Raman spectroscopy is performed on the four a-SiC films, the peaks on the spectrum correspond to Si-Si and C-C bonds are found around 960 and 1455 cm$^{-1}$. Ratios between heights of two peaks for each a-SiC film are shown, the a-SiCR2 film shows higher C-C bond proportion than the a-SiCR4 film, which may explain the former has higher ultimate tensile strength. The spectrum are spaced between each other for clear view.  (d) Zoom in of the Raman spectrum correspond to the Si-C bonds in (c) \cite{dey2020tailoring}, one can notice that the spectrum of a-SiCR2/R3/170 are similar to each other, while the one of a-SiCR4 looks a bit different from the others.}
\label{AMreply2}
\end{figure*}

\section*{Supporting Information \textbf{(I)}: Fabrication process}

Low pressure chemical vapor deposition (LPCVD) non-stoichiometric a-SiC films were used in this paper, deposited with different gas flow ratios (GFR) between SiH$_2$Cl$_2$ and 5\% C$_2$H$_2$ in H$_2$ (GFR=2,3,4), at various deposition pressures (170/600 mTorr), and on both silicon and fused silica substrates (\textbf{Table \ref{table:1}}). The variation in deposition parameters allows us to systematically characterize the mechanical properties of LPCVD a-SiC. All a-SiC films were deposited at a temperature 760C for the same period of time (3 hours 20 minutes) to avoid film property differences caused by thermal effects and to ensure that the SiC films were composited of amorphous form instead of poly-crystalline form \cite{MoranaPhD2015, nakao2008mechanical}.

After LPCVD a-SiC deposition, the wafers are diced into smaller chips. The chips are then exposed to electron beam lithography to create desired patterns on the e-beam resist coated on top. Subsequently, these patterns are transferred into the a-SiC films using CHF$_3$ anisotropic plasma etching. Next, the patterned chips are cleaned with dimethylformamide and Piranha solution, followed by the undercut of the silicon substrate or fused silica substrate using cryogenic SF$_6$ isotropic plasma etching or vapor hydrofluoric acid. Finally, the designed a-SiC nanomechanical resonators are fabricated.

\begin{figure*}[h]
\centering
\includegraphics[width=1\textwidth]{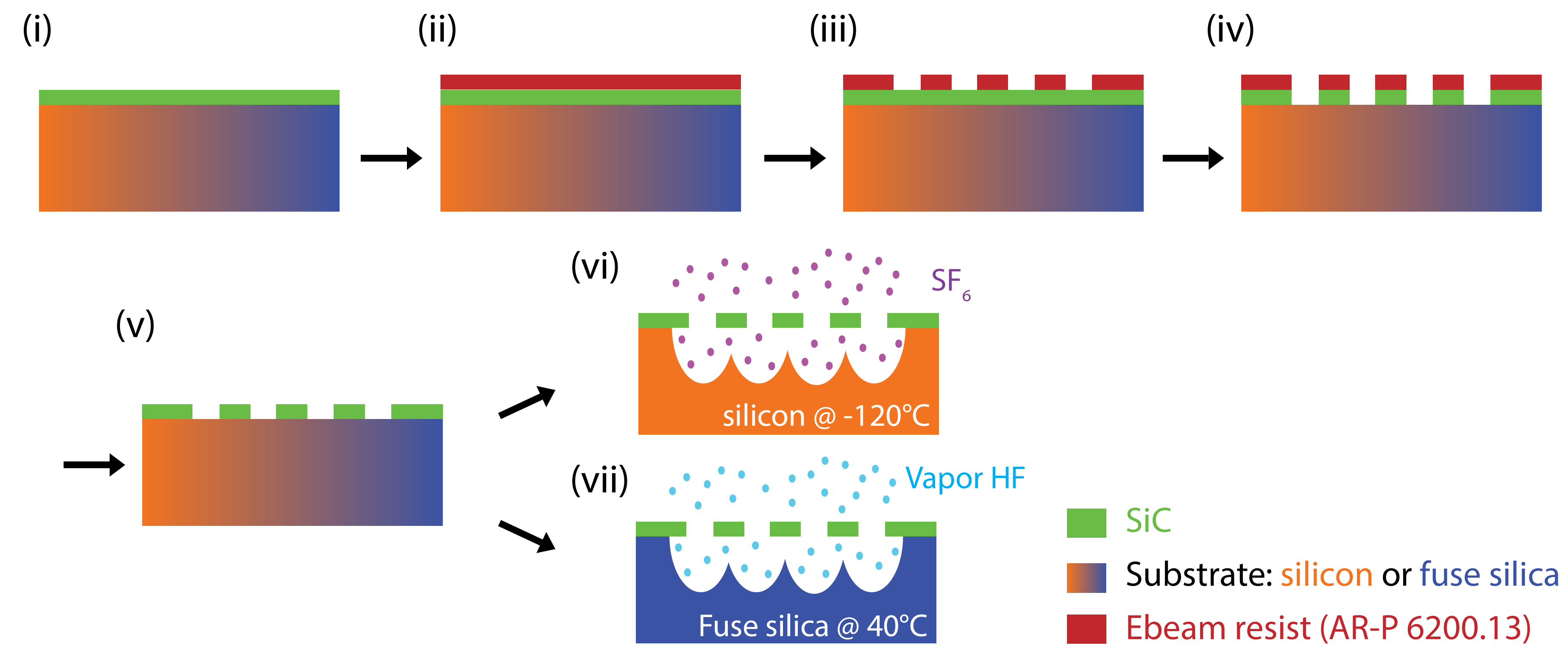}
\caption{Fabrication process flow to fabricate a-SiC (green) nanomechanical resonators on both silicon (orange) and fused silica substrates (blue). (i) Start with a die of the wafer. (ii) Spin coat ebeam resist AR-P 6200.13 at spin speed 1500 rpm, then bake at 150 $^{\circ}$C for 3 mins. (iii) Ebeam pattern of the desired pattern, afterwards developing the pattern by sequently immersing the die in Pentyl Acetate, MiBK:IPA=1:1, IPA solutions for 1 min. (iv) Transfer the pattern into the a-SiC film with reactive ion etching (Sentech Etchlab 200), with CHF$_3$/Ar plasma at the power 60 watts. (v) Remove the ebeam resist with hot Dimethylformamide solution in a supersonic bath, following by Piranha and diluted hydrofluoric acid cleaning. (vi) Undercut the silicon substrate to suspend a-SiC resonators with cryogenic (-120$^{\circ}$C) SF$_6$ plasma etching. (vii) Undercut the fused silica substrate to suspend a-SiC resonators with vapor hydrofluoric acid (at 35 to 60$^{\circ}$C).}
\end{figure*}

\newpage

\section*{Supporting Information \textbf{(J)}: Ringdown Measurement with Homodyne Detection}

We use a balanced homodyne interferometer for performing ringdown experiments on a-SiC nanoemchanical resonators. As shown in Figure \ref{fig:interferometry}, the a-SiC nanomechanical resonator (green) on top on the substrate (brown) is placed in an ultra-high vacuum (UHV) chamber under a pressure lower than $10^{-8}$ mbar. This avoids mechanical losses due to gas damping. Ringdown measurements are performed via a piezoelectric actuator which resonantly drives the corrugated nanostrings. After reaching maximal amplitude, the drive is stopped to observe the rate at which mechanical energy is dissipated from the nanostrings. The vibration amplitude of the resonator is measured optically with a fiber coupled infrared laser (1550 nm). The power of the laser is divided into two parts, 90\% of it is used for interference reference (local oscillator), while the other 10\% terminates with a lensed fiber shines on the resonator. The reflected light from the resonator then compares its phase to the one from the local oscillator using the balanced homodyne measurement setup, with which the amplitude of the resonator is measured. 

\begin{figure*}[h]
\centering
\includegraphics[width=1\textwidth]{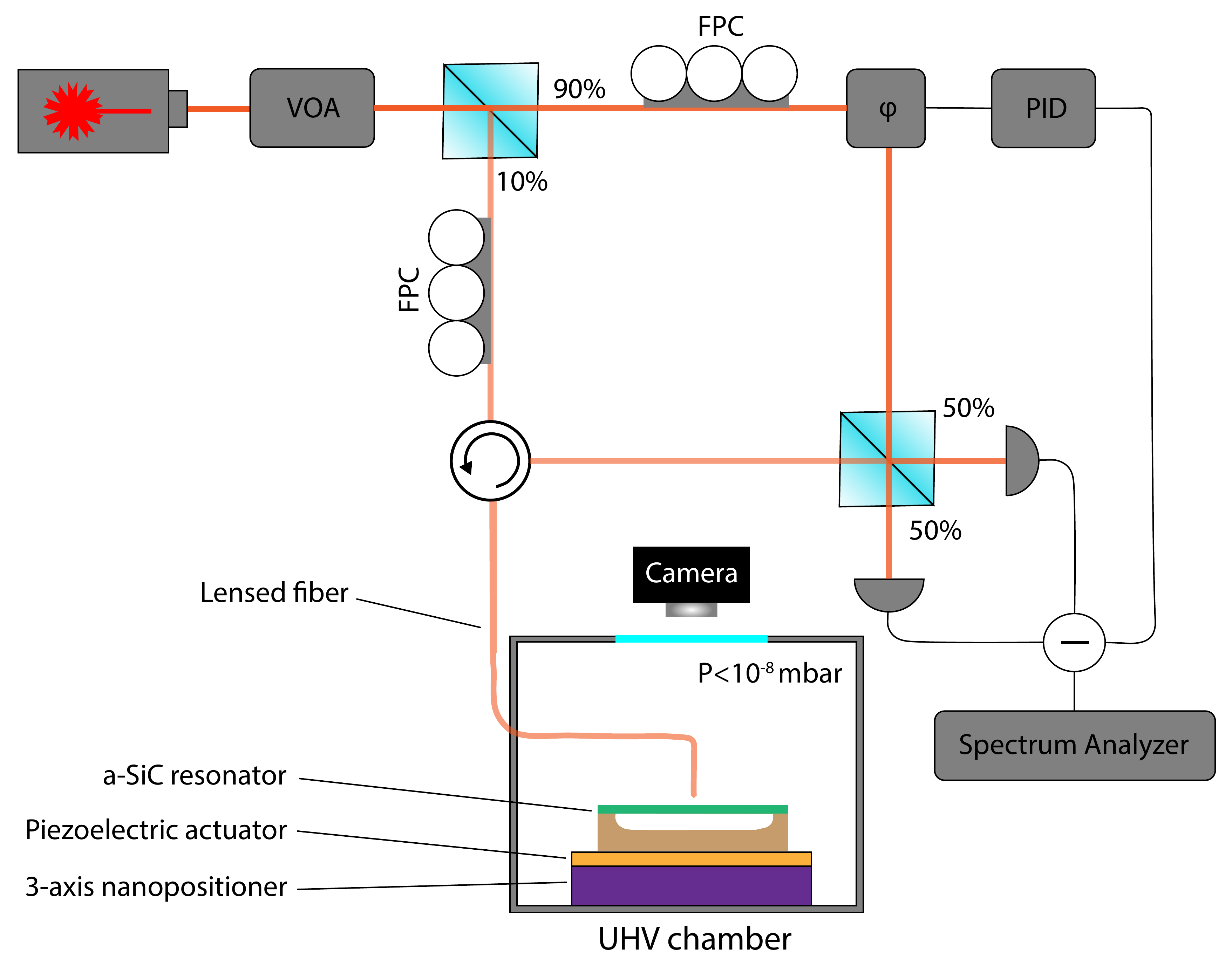}
\caption{Schematic of the balanced homodyne interferometer used for performing ringdown experiments. The nanomechanical resonator, placed in an UHV chamber, is resonantly driven by a piezoelectric actuator, and its motion is measured with the interferometer. Abbreviations: VOA, Variable Optical Attenuator; FPC, Fiber Polarization Controller; $\varphi$: Phase shifter; PID: Proportional–Integral–Derivative controller}
\label{fig:interferometry}
\end{figure*}

\end{widetext}


\begin{thebibliography}{0}%
\makeatletter
\providecommand \@ifxundefined [1]{%
 \@ifx{#1\undefined}
}%
\providecommand \@ifnum [1]{%
 \ifnum #1\expandafter \@firstoftwo
 \else \expandafter \@secondoftwo
 \fi
}%
\providecommand \@ifx [1]{%
 \ifx #1\expandafter \@firstoftwo
 \else \expandafter \@secondoftwo
 \fi
}%
\providecommand \natexlab [1]{#1}%
\providecommand \enquote  [1]{``#1''}%
\providecommand \bibnamefont  [1]{#1}%
\providecommand \bibfnamefont [1]{#1}%
\providecommand \citenamefont [1]{#1}%
\providecommand \href@noop [0]{\@secondoftwo}%
\providecommand \href [0]{\begingroup \@sanitize@url \@href}%
\providecommand \@href[1]{\@@startlink{#1}\@@href}%
\providecommand \@@href[1]{\endgroup#1\@@endlink}%
\providecommand \@sanitize@url [0]{\catcode `\\12\catcode `\$12\catcode `\&12\catcode `\#12\catcode `\^12\catcode `\_12\catcode `\%12\relax}%
\providecommand \@@startlink[1]{}%
\providecommand \@@endlink[0]{}%
\providecommand \url  [0]{\begingroup\@sanitize@url \@url }%
\providecommand \@url [1]{\endgroup\@href {#1}{\urlprefix }}%
\providecommand \urlprefix  [0]{URL }%
\providecommand \Eprint [0]{\href }%
\providecommand \doibase [0]{https://doi.org/}%
\providecommand \selectlanguage [0]{\@gobble}%
\providecommand \bibinfo  [0]{\@secondoftwo}%
\providecommand \bibfield  [0]{\@secondoftwo}%
\providecommand \translation [1]{[#1]}%
\providecommand \BibitemOpen [0]{}%
\providecommand \bibitemStop [0]{}%
\providecommand \bibitemNoStop [0]{.\EOS\space}%
\providecommand \EOS [0]{\spacefactor3000\relax}%
\providecommand \BibitemShut  [1]{\csname bibitem#1\endcsname}%
\let\auto@bib@innerbib\@empty
\end{thebibliography}%


\begin{thebibliography}{10}
\providecommand{\url}[1]{\texttt{#1}}
\providecommand{\urlprefix}{URL }

\bibitem{Krause2012}
A.~G. Krause, M.~Winger, T.~D. Blasius, Q.~Lin, O.~Painter,
\newblock \emph{Nature Photonics} \textbf{2012}, \emph{6}, 11 768.

\bibitem{Pratt2021}
J.~R. Pratt, A.~R. Agrawal, C.~A. Condos, C.~M. Pluchar, S.~Schlamminger, D.~J. Wilson,
\newblock \emph{Physical Review X} \textbf{2023}, \emph{13}, 1 011018.

\bibitem{Manzaneque2022}
T.~Manzaneque, M.~K. Ghatkesar, F.~Alijani, M.~Xu, R.~A. Norte, P.~G. Steeneken,
\newblock \emph{arXiv preprint arXiv:2205.11903} \textbf{2022}.

\bibitem{Sanchez2012}
D.~Garcia-Sanchez, K.~Y. Fong, H.~Bhaskaran, S.~Lamoreaux, H.~X. Tang,
\newblock \emph{Physical review letters} \textbf{2012}, \emph{109}, 2 027202.

\bibitem{Pate2020}
J.~M. Pate, M.~Goryachev, R.~Y. Chiao, J.~E. Sharping, M.~E. Tobar,
\newblock \emph{Nature Physics} \textbf{2020}, \emph{16}, 11 1117.

\bibitem{halg2021membrane}
D.~H{\"a}lg, T.~Gisler, Y.~Tsaturyan, L.~Catalini, U.~Grob, M.-D. Krass, M.~H{\'e}ritier, H.~Mattiat, A.-K. Thamm, R.~Schirhagl, et~al.,
\newblock \emph{Physical Review Applied} \textbf{2021}, \emph{15}, 2 L021001.

\bibitem{reinhardt2016}
C.~Reinhardt, T.~M{\"u}ller, A.~Bourassa, J.~C. Sankey,
\newblock \emph{Physical Review X} \textbf{2016}, \emph{6}, 2 021001.

\bibitem{Verbridge2006}
S.~S. Verbridge, J.~M. Parpia, R.~B. Reichenbach, L.~M. Bellan, H.~G. Craighead,
\newblock \emph{Journal of Applied Physics} \textbf{2006}, \emph{99}, 12 124304.

\bibitem{Norte2016}
R.~A. Norte, J.~P. Moura, S.~Gr{\"o}blacher,
\newblock \emph{Physical review letters} \textbf{2016}, \emph{116}, 14 147202.

\bibitem{Ghadimi2018}
A.~H. Ghadimi, S.~A. Fedorov, N.~J. Engelsen, M.~J. Bereyhi, R.~Schilling, D.~J. Wilson, T.~J. Kippenberg,
\newblock \emph{Science} \textbf{2018}, \emph{360}, 6390 764.

\bibitem{Fedorov2020}
S.~A. Fedorov, A.~Beccari, N.~J. Engelsen, T.~J. Kippenberg,
\newblock \emph{Physical Review Letters} \textbf{2020}, \emph{124}, 2 025502.



\bibitem{Bereyhi2022}
M.~J. Bereyhi, A.~Arabmoheghi, A.~Beccari, S.~A. Fedorov, G.~Huang, T.~J. Kippenberg, N.~J. Engelsen,
\newblock \emph{Physical Review X} \textbf{2022}, \emph{12}, 2 021036.

\bibitem{Beccari2022}
A.~Beccari, D.~A. Visani, S.~A. Fedorov, M.~J. Bereyhi, V.~Boureau, N.~J. Engelsen, T.~J. Kippenberg,
\newblock \emph{Nature Physics} \textbf{2022}, \emph{18}, 4 436.

\bibitem{Romero2020}
E.~Romero, V.~M. Valenzuela, A.~R. Kermany, L.~Sementilli, F.~Iacopi, W.~P. Bowen,
\newblock \emph{Physical Review Applied} \textbf{2020}, \emph{13}, 4 044007.

\bibitem{Klas2022}
Y.~S. Kla{\ss},
\newblock Ph.D. thesis, Technische Universit{\"a}t M{\"u}nchen, \textbf{2022}.

\bibitem{Manjeshwar2022}
S.~K. Manjeshwar, A.~Ciers, F.~Hellman, J.~Bl{\"a}sing, A.~Strittmater, W.~Wieczorek,
\newblock \emph{arXiv preprint arXiv:2211.12469} \textbf{2022}.

\bibitem{gu2013microstructure}
X.~W. Gu, Z.~Wu, Y.-W. Zhang, D.~J. Srolovitz, J.~R. Greer,
\newblock \emph{Nano letters} \textbf{2013}, \emph{13}, 11 5703.

\bibitem{zhao2022controlling}
X.~Zhao, B.~Mao, M.~Liu, J.~Cao, S.~J. Haigh, D.~G. Papageorgiou, Z.~Li, R.~J. Young,
\newblock \emph{Advanced Functional Materials} \textbf{2022}, \emph{32}, 42 2202373.

\bibitem{Bereyhi2019}
M.~J. Bereyhi, A.~Beccari, S.~A. Fedorov, A.~H. Ghadimi, R.~Schilling, D.~J. Wilson, N.~J. Engelsen, T.~J. Kippenberg,
\newblock \emph{Nano letters} \textbf{2019}, \emph{19}, 4 2329.

\bibitem{Kwon2015}
G.~Kwon, H.-H. Jo, S.~Lim, C.~Shin, H.-H. Jin, J.~Kwon, G.-M. Sun,
\newblock \emph{Journal of materials science} \textbf{2015}, \emph{50} 8104.

\bibitem{Rasool2013}
H.~I. Rasool, C.~Ophus, W.~S. Klug, A.~Zettl, J.~K. Gimzewski,
\newblock \emph{Nature communications} \textbf{2013}, \emph{4}, 1 2811.

\bibitem{zhang2015fracture}
T.~Zhang, X.~Li, H.~Gao,
\newblock \emph{International Journal of Fracture} \textbf{2015}, \emph{196} 1.

\bibitem{Goldsche2018}
M.~Goldsche, J.~Sonntag, T.~Khodkov, G.~J. Verbiest, S.~Reichardt, C.~Neumann, T.~Ouaj, N.~von~den Driesch, D.~Buca, C.~Stampfer,
\newblock \emph{Nano letters} \textbf{2018}, \emph{18}, 3 1707.

\bibitem{lee2008measurement}
C.~Lee, X.~Wei, J.~W. Kysar, J.~Hone,
\newblock \emph{science} \textbf{2008}, \emph{321}, 5887 385.

\bibitem{wang2012effect}
M.~Wang, C.~Yan, L.~Ma, N.~Hu, M.~Chen,
\newblock \emph{Computational Materials Science} \textbf{2012}, \emph{54} 236.

\bibitem{xu2018enhancing}
J.~Xu, G.~Yuan, Q.~Zhu, J.~Wang, S.~Tang, L.~Gao,
\newblock \emph{ACS nano} \textbf{2018}, \emph{12}, 5 4529.

\bibitem{qu2012metallic}
R.~Qu, M.~Calin, J.~Eckert, Z.~Zhang,
\newblock \emph{Scripta Materialia} \textbf{2012}, \emph{66}, 10 733.

\bibitem{jiang2021structures}
H.~Jiang, T.~Shang, H.~Xian, B.~Sun, Q.~Zhang, Q.~Yu, H.~Bai, L.~Gu, W.~Wang,
\newblock \emph{Small Structures} \textbf{2021}, \emph{2}, 2 2000057.

\bibitem{greer1995metallic}
A.~L. Greer,
\newblock \emph{Science} \textbf{1995}, \emph{267}, 5206 1947.

\bibitem{telford2004case}
M.~Telford,
\newblock \emph{Materials today} \textbf{2004}, \emph{7}, 3 36.

\bibitem{Wijesundara2011}
M.~Wijesundara, R.~Azevedo,
\newblock \emph{Silicon carbide microsystems for harsh environments}, volume~22,
\newblock Springer Science \& Business Media, \textbf{2011}.

\bibitem{Gerhardt2011}
R.~Gerhardt,
\newblock \emph{Properties and applications of silicon carbide},
\newblock BoD--Books on Demand, \textbf{2011}.

\bibitem{Kimoto2014}
T.~Kimoto, J.~A. Cooper,
\newblock \emph{Fundamentals of Silicon Carbide Technology: Growth, Characterization, Devices and Applications},
\newblock Wiley-IEEE Press, \textbf{2014}.

\bibitem{Morana2012}
B.~Morana, G.~Pandraud, J.~Creemer, P.~Sarro,
\newblock \emph{Materials Chemistry and Physics} \textbf{2013}, \emph{139}, 2-3 654.

\bibitem{Iliescu2012}
C.~Iliescu, D.~P. Poenar,
\newblock In \emph{Physics and Technology of Silicon Carbide Devices}. IntechOpen, \textbf{2012}.

\bibitem{Blum1999}
T.~Blum, B.~Dresler, M.~Hoffmann, et~al.,
\newblock \emph{Surface and Coatings Technology} \textbf{1999}, \emph{116} 1024.

\bibitem{Jiang1999}
L.~Jiang, X.~Chen, X.~Wang, L.~Xu, F.~Stubhan, K.-H. Merkel,
\newblock \emph{Thin Solid Films} \textbf{1999}, \emph{352}, 1-2 97.

\bibitem{Castelletto2020}
S.~Castelletto, A.~Boretti,
\newblock \emph{Journal of Physics: Photonics} \textbf{2020}, \emph{2}, 2 022001.

\bibitem{Kohler2021}
M.~K{\"o}hler, M.~Pomaska, P.~Procel, R.~Santbergen, A.~Zamchiy, B.~Macco, A.~Lambertz, W.~Duan, P.~Cao, B.~Klingebiel, et~al.,
\newblock \emph{Nature Energy} \textbf{2021}, \emph{6}, 5 529.

\bibitem{Sementilli2022}
L.~Sementilli, E.~Romero, W.~P. Bowen,
\newblock \emph{Advanced Functional Materials} \textbf{2022}, \emph{32}, 3 2105247.

\bibitem{nguyen2017nanopore}
T.-K. Nguyen, H.-P. Phan, H.~Kamble, R.~Vadivelu, T.~Dinh, A.~Iacopi, G.~Walker, L.~Hold, N.-T. Nguyen, D.~V. Dao,
\newblock \emph{ACS applied materials \& interfaces} \textbf{2017}, \emph{9}, 48 41641.

\bibitem{Atwater2018}
H.~A. Atwater, A.~R. Davoyan, O.~Ilic, D.~Jariwala, M.~C. Sherrott, C.~M. Went, W.~S. Whitney, J.~Wong,
\newblock \emph{Nature materials} \textbf{2018}, \emph{17}, 10 861.

\bibitem{MoranaPhD2015}
B.~Morana,
\newblock Ph.D. thesis, TU Delft, \textbf{2015}.

\bibitem{Chu2009}
J.~Chu, D.~Zhang,
\newblock \emph{Journal of Micromechanics and Microengineering} \textbf{2009}, \emph{19}, 9 095020.

\bibitem{Imran2018}
M.~Imran, M.~Mahendran, P.~Keerthan,
\newblock \emph{Journal of Constructional Steel Research} \textbf{2018}, \emph{143} 131.

\bibitem{Zhou2004}
Y.~Zhou, Y.~Wang, P.~Mallick,
\newblock \emph{Materials Science and Engineering: A} \textbf{2004}, \emph{381}, 1-2 355.

\bibitem{Chen2004}
K.-S. Chen, A.~Ayon, S.~M. Spearing,
\newblock \emph{Journal of the American Ceramic Society} \textbf{2000}, \emph{83}, 6 1476.

\bibitem{Cui2019}
J.~Cui, Z.~Zhang, H.~Jiang, D.~Liu, L.~Zou, X.~Guo, Y.~Lu, I.~P. Parkin, D.~Guo,
\newblock \emph{ACS nano} \textbf{2019}, \emph{13}, 7 7483.

\bibitem{Shuman2007}
D.~J. Shuman, A.~L. Costa, M.~S. Andrade,
\newblock \emph{Materials characterization} \textbf{2007}, \emph{58}, 4 380.

\bibitem{Kim2003}
J.-H. Kim, S.-C. Yeon, Y.-K. Jeon, J.-G. Kim, Y.-H. Kim,
\newblock \emph{Sensors and Actuators A: Physical} \textbf{2003}, \emph{108}, 1-3 20.

\bibitem{Klass2022}
Y.~S. Kla{\ss}, J.~Doster, M.~B{\"u}ckle, R.~Braive, E.~M. Weig,
\newblock \emph{Applied Physics Letters} \textbf{2022}, \emph{121}, 8 083501.

\bibitem{Barboni2018}
L.~Barboni, G.~Gillich, C.~Chioncel, C.~Hamat, I.~Mituletu,
\newblock In \emph{IOP Conference Series: Materials Science and Engineering}, volume 416. IOP Publishing, \textbf{2018} 012063.

\bibitem{Chirikov2020}
V.~A. Chirikov, D.~M. Dimitrov, Y.~S. Boyadjiev,
\newblock \emph{Procedia Manufacturing} \textbf{2020}, \emph{46} 87.

\bibitem{Chen2000}
S.~Chen,
\newblock \emph{Ultrasonics} \textbf{2000}, \emph{38}, 1-8 206.

\bibitem{Villanueva2014}
L.~G. Villanueva, S.~Schmid,
\newblock \emph{Physical review letters} \textbf{2014}, \emph{113}, 22 227201.

\bibitem{Wang2017}
S.~Wang, Z.~Shan, H.~Huang,
\newblock \emph{Advanced Science} \textbf{2017}, \emph{4}, 4 1600332.

\bibitem{banerjee2018ultralarge}
A.~Banerjee, D.~Bernoulli, H.~Zhang, M.-F. Yuen, J.~Liu, J.~Dong, F.~Ding, J.~Lu, M.~Dao, W.~Zhang, et~al.,
\newblock \emph{Science} \textbf{2018}, \emph{360}, 6386 300.

\bibitem{Shafikov2021}
A.~Shafikov, B.~Schurink, R.~W. van~de Kruijs, J.~Benschop, W.~Van~den Beld, Z.~S. Houweling, F.~Bijkerk,
\newblock \emph{Sensors and Actuators A: Physical} \textbf{2021}, \emph{317} 112456.

\bibitem{Demetriou2011}
M.~D. Demetriou, M.~E. Launey, G.~Garrett, J.~P. Schramm, D.~C. Hofmann, W.~L. Johnson, R.~O. Ritchie,
\newblock \emph{Nature materials} \textbf{2011}, \emph{10}, 2 123.

\bibitem{Zhang2022}
S.~Zhang, Z.~Li, K.~Luo, J.~He, Y.~Gao, A.~V. Soldatov, V.~Benavides, K.~Shi, A.~Nie, B.~Zhang, et~al.,
\newblock \emph{National Science Review} \textbf{2022}, \emph{9}, 1 nwab140.

\bibitem{namazu2023mechanical}
T.~Namazu,
\newblock \emph{IEEJ Transactions on Electrical and Electronic Engineering} \textbf{2023}, \emph{18}, 3 308.

\bibitem{grunenberg2001intrinsic}
J.~Grunenberg,
\newblock \emph{Angewandte Chemie International Edition} \textbf{2001}, \emph{40}, 21 4027.

\bibitem{villanueva2014evidence}
L.~G. Villanueva, S.~Schmid,
\newblock \emph{Physical review letters} \textbf{2014}, \emph{113}, 22 227201.

\bibitem{Fedorov2019}
S.~A. Fedorov, N.~J. Engelsen, A.~H. Ghadimi, M.~J. Bereyhi, R.~Schilling, D.~J. Wilson, T.~J. Kippenberg,
\newblock \emph{Physical Review B} \textbf{2019}, \emph{99}, 5 054107.

\bibitem{bagci2014optical}
T.~Bagci, A.~Simonsen, S.~Schmid, L.~G. Villanueva, E.~Zeuthen, J.~Appel, J.~M. Taylor, A.~S{\o}rensen, K.~Usami, A.~Schliesser, et~al.,
\newblock \emph{Nature} \textbf{2014}, \emph{507}, 7490 81.

\bibitem{teufel2009nanomechanical}
J.~D. Teufel, T.~Donner, M.~Castellanos-Beltran, J.~W. Harlow, K.~W. Lehnert,
\newblock \emph{Nature nanotechnology} \textbf{2009}, \emph{4}, 12 820.

\bibitem{Guo2019}
J.~Guo, R.~Norte, S.~Gr{\"o}blacher,
\newblock \emph{Physical review letters} \textbf{2019}, \emph{123}, 22 223602.

\bibitem{Gonzalez1994}
G.~I. Gonz{\'a}lez, P.~R. Saulson,
\newblock \emph{The Journal of the Acoustical Society of America} \textbf{1994}, \emph{96}, 1 207.

\bibitem{Tsaturyan2017}
Y.~Tsaturyan, A.~Barg, E.~S. Polzik, A.~Schliesser,
\newblock \emph{Nature nanotechnology} \textbf{2017}, \emph{12}, 8 776.

\bibitem{Li2023}
Z.~Li, M.~Xu, R.~A. Norte, A.~M. Arag{\'o}n, F.~Van~Keulen, F.~Alijani, P.~G. Steeneken,
\newblock \emph{Applied Physics Letters} \textbf{2023}, \emph{122}, 1 013501.

\bibitem{Hoj2022}
D.~H{\o}j, U.~B. Hoff, U.~L. Andersen,
\newblock \emph{arXiv preprint arXiv:2207.06703} \textbf{2022}.

\bibitem{Sadeghi2021}
P.~Sadeghi,
\newblock Ph.D. thesis, Wien, \textbf{2021}.

\bibitem{Gisler2022}
T.~Gisler, M.~Helal, D.~Sabonis, U.~Grob, M.~H{\'e}ritier, C.~L. Degen, A.~H. Ghadimi, A.~Eichler,
\newblock \emph{Physical Review Letters} \textbf{2022}, \emph{129}, 10 104301.

\bibitem{nakao2008mechanical}
S.~Nakao, T.~Ando, L.~Chen, M.~Mehregany, K.~Sato,
\newblock In \emph{2008 IEEE 21st International Conference on Micro Electro Mechanical Systems}. IEEE, \textbf{2008} 447--450.


\bibitem{Shin2021}
D.~Shin, A.~Cupertino, M.~H. de~Jong, P.~G. Steeneken, M.~A. Bessa, R.~A. Norte,
\newblock \emph{Advanced Materials} \textbf{2022}, \emph{34}, 3 2106248.



\bibitem{Wilson2009}
D.~J. Wilson, C.~A. Regal, S.~B. Papp, H.~Kimble,
\newblock \emph{Physical review letters} \textbf{2009}, \emph{103}, 20 207204.

\bibitem{Buckle2021}
M.~B{\"u}ckle, Y.~S. Kla{\ss}, F.~B. N{\"a}gele, R.~Braive, E.~M. Weig,
\newblock \emph{Physical Review Applied} \textbf{2021}, \emph{15}, 3 034063.

\bibitem{macho2015oscillations}
E.~Macho-Stadler, M.~Elejalde-Garc{\'\i}a, R.~Llanos-V{\'a}zquez,
\newblock \emph{European Journal of Physics} \textbf{2015}, \emph{36}, 5 055007.

\bibitem{steeneken2021dynamics}
P.~G. Steeneken, R.~J. Dolleman, D.~Davidovikj, F.~Alijani, H.~S. Van~der Zant,
\newblock \emph{2D Materials} \textbf{2021}, \emph{8}, 4 042001.

\bibitem{castellanos2013single}
A.~Castellanos-Gomez, R.~van Leeuwen, M.~Buscema, H.~S. van~der Zant, G.~A. Steele, W.~J. Venstra,
\newblock \emph{Advanced Materials} \textbf{2013}, \emph{25}, 46 6719.

\bibitem{duvigneau2016vibration}
F.~Duvigneau, S.~Koch, R.~Orszulik, E.~Woschke, U.~Gabbert,
\newblock \emph{Tech Mech} \textbf{2016}, \emph{36}, 3 180.


\bibitem{Hoj2021}
D.~H{\o}j, F.~Wang, W.~Gao, U.~B. Hoff, O.~Sigmund, U.~L. Andersen,
\newblock \emph{Nature communications} \textbf{2021}, \emph{12}, 1 5766.


\bibitem{pereira2009situ}
J.~Pereira, L.~E. Pichon, R.~Dussart, C.~Cardinaud, C.~Y. Duluard, E.-H. Oubensaid, P.~Lefaucheux, M.~Boufnichel, P.~Ranson,
\newblock \emph{Applied Physics Letters} \textbf{2009}, \emph{94}, 7.

\bibitem{kaur2015selective}
A.~Kaur, P.~Chahal, T.~Hogan,
\newblock \emph{IEEE Electron Device Letters} \textbf{2015}, \emph{37}, 2 142.

\bibitem{mckenna2011synthesis}
J.~McKenna, J.~Patel, S.~Mitra, N.~Soin, V.~{\v{S}}vr{\v{c}}ek, P.~Maguire, D.~Mariotti,
\newblock \emph{The European Physical Journal-Applied Physics} \textbf{2011}, \emph{56}, 2 24020.

\bibitem{yu2019third}
X.~Yu, B.~Ding, H.~Lu, Y.~Huo, Q.~Peng, X.~Xiu, C.~Zhang, C.~Yang, S.~Jiang, B.~Man, et~al.,
\newblock \emph{Journal of Alloys and Compounds} \textbf{2019}, \emph{794} 518.


\bibitem{dey2020tailoring}
P.~P. Dey, A.~Khare,
\newblock \emph{SN Applied Sciences} \textbf{2020}, \emph{2} 1.




\end{thebibliography}
\end{document}